\documentclass[pra,twocolumn,preprintnumbers,superscriptaddress]{revtex4}

\usepackage{amsmath,amsfonts,amssymb,bbm,slashed}
\usepackage[english]{babel}
\usepackage{graphicx}
\usepackage{color}
\usepackage[utf8]{inputenc}
\usepackage[T1]{fontenc}
\usepackage{tensor}
\usepackage{ dsfont }
\usepackage{mathtools}
\usepackage{float}
\usepackage{wrapfig}
\usepackage{stackengine}

\begin{document}
	
	\title{Dirac equation on a square waveguide lattice with site-dependent coupling strengths and the gravitational Aharonov-Bohm effect}
	
	\author{Christian Koke}
	\email{christian.koke@campus.lmu.de}
	\affiliation{Arnold Sommerfeld Center, Ludwig Maximilian University of Munich, Munich 80333, Germany}
	
	\author{Changsuk Noh}
	\email{cnoh@knu.ac.kr}
	\affiliation{Department of Physics, Kyungpook National University,  Daegu 41566, Korea}
	
	\author{Dimitris G. Angelakis}
	\affiliation{Centre for Quantum Technologies, National University of Singapore, Singapore 117542}
	\affiliation{School of Electrical and Computer Engineering, Technical University of Crete, Chania 73100, Greece}

	\begin{abstract}
		The main objective of this work is to present a theoretical proposal for an implementation of the $(2 + 1)$-dimensional Dirac equation in classical gravitational and electromagnetic backgrounds in a two-dimensional	waveguide array. For this, a framework for achieving site-dependent effective coupling constants in two-dimensional
		waveguide arrays is developed. Implementability of the Dirac equation under the proposed scheme puts minor demands on gauge and spacetime backgrounds; however, a wide array of physical spacetimes, such
		as all vacuum and static solutions, prove to be implementable. As an interesting and instructive example, we discuss a tabletop realization of the gravitational Aharonov-Bohm effect: After devising a thought experiment in which signatures of the gravitational Aharonov-Bohm effect could be detected, we briefly discuss how the analogue of such a setting can in principle be implemented using the proposed waveguide setup.
	\end{abstract}

	\maketitle
	\section{introduction}
	Physical effects stemming from global geometric or topological properties of a system have long been investigated in physics. A premier example of such an effect is the celebrated Aharonov-Bohm effect in  quantum mechanics \cite{Aharonov}. 
	Similar phenomena can occur in $2+1$ dimensional gravity, where localised sources do not influence the curvature of the manifold beyond their point of localisation, but may effect the global geometry \cite{Deser}.
	In this work we propose a classical optical simulation of a gravitational analogue of the Aharonov-Bohm effect for Fermions in $2+1$ dimensions \cite{Bezerra, Ford, Vilenkin0, Burges}.
	
	There exists a multitude of  proposals and experimental demonstrations of interesting physics in coupled waveguide arrays
	\cite{Peschel, Morandotti, Lahini, Longhi11, Longhi12, Crespi, Keil, Rodriguez13, Rodriguez14, LeeAngelakis14, Marini14, RaiAngelakis15, Keil15}.
	Optical simulations of the $(1+1)$-dimensional Dirac equation in binary waveguide arrays have long been proposed \cite{Longhi10a,Longhi10b} and experimentally demonstrated \cite{Dreisow10,Dreisow12}.
	Simulations of Dirac dynamics in curved spacetime have also been investigated. 
	For approaches in graphene see Refs. \cite{Boada,Setare}; investigations using cold atoms, e.g. Refs. \cite{Kosior, Celi}, are also conducted.
	For the general simulation of the $(1+1)$-dimensional Dirac equation in curved spacetime in coupled waveguide arrays, see \cite{Koke}.
	We now expand on this work by elaborating on how the $(2+1)$-dimensional Dirac equation may be simulated in coupled waveguide arrays and then show how one can incorporate classical gravitational and electromagnetic background fields.
	
	This article is organized as follows: After discussing the types of metrics and gauge connections we will consider, we recall the basics of the $(2+1)$-dimensional Dirac equation coupled to background fields in section \ref{generalDiracsec}. Subsequently, generalizing an idea proposed in Ref. \cite{Efremidis}, we show how one can achieve site dependent effective coupling constants in two-dimensional coupled waveguide arrays in section \ref{photlat}. Section \ref{impl} then explains how one can use this to implement the Dirac equation in square lattice waveguide arrays. Finally, in section \ref{experiment} we develop a scenario in which the 'phase shift' of the gravitational Aharonov-Bohm effect can be observed and discuss its potential realization in a coupled waveguide array.	
	\section{The minimally coupled Dirac equation in curved spacetime }\label{generalDiracsec}
	In what follows, we shall always assume a global manifold chart.
	\subsection{General considerations}
	Our starting point is the coordinate expression of the minimally coupled Dirac equation in curved spacetime:
	\begin{align}
		\left[ i\gamma^\mu\nabla_\mu -m\right]\psi(x) = 0
	\end{align}
	Here we are using the vielbein $e\indices{^\mu_a}(x)$,
	to transform the local gamma matrices $\gamma^a$ according to
	\begin{equation}
		\gamma^\mu(x) = e\indices{^\mu_a}(x)\gamma^a.
	\end{equation}
	The covariant derivative is given by
	\begin{equation}
		\nabla_\nu = \partial_\nu + \Omega_\nu + i A_{\nu},
	\end{equation}
	with a  gauge connection  $ A_{\nu}$ and where  $\Omega_\nu$ arises from the spin connection \cite{Birell}.
	For a brief introduction to the Dirac equation in curved spacetime, the reader might want to consider the Appendix.
	%
	For reasons that will become apparent shortly, we assume that the metric is of the form
	\begin{equation}\label{prototype}
		ds^2= e^{2\Phi(x,y,t)}dt^2-e^{2\Psi(x,y,t)}(dx^2+dy^2),
	\end{equation}
	and that 
	\begin{equation}\label{timeind}
		\Phi(x,y,t) - \Psi(x,y,t) = \rho(x,y),
	\end{equation}
	where $\rho$ is independent of time.
	These restrictions still allow for all vacuum solutions, since, as argued in Appendix, vacuum solutions can be taken to be conformally flat and hence we may even choose $\Phi = \Psi$.  Furthermore we note that any static spacetime may be represented in this form \cite{Boada}. 
	For the gauge connection, we demand that the vector part may be gauged away, i.e.~that we may write
	\begin{equation}
		\begin{split}
			A_{\mu}d^{\mu}x=& \phi(x,y,t) dt + \\
			&\partial_x\Lambda(x,y,t) dx + \partial_y \Lambda(x,y,t) dy.
		\end{split}
	\end{equation}
	%
	
	Under these assumptions, after choosing $\gamma^0 = \sigma_z$, $\gamma^1=  - i\sigma_x$ and $\gamma^2=  i\sigma_y$, 
	and first neglecting the gauge part, the Dirac equation can be written as 
	\begin{equation}
		\begin{split}
			&i\left(\partial_t+\Psi_t\right)\psi= -i\sigma_ye^{\Phi-\Psi}\left(\partial_x + \frac{1}{2}\Psi_x +\frac{1}{2}\Phi_x \right)\psi\\
			&-i\sigma_xe^{\Phi-\Psi}\left(\partial_y + \frac{1}{2}\Psi_y +\frac{1}{2}\Phi_y \right)\psi  +\sigma_ze^{\Phi} m \psi.
		\end{split}
	\end{equation}
	See Appendix \ref{Diraccoordgeneral} for calculations leading up to this.
	Recalling that $\Phi-\Psi = \rho$ and reinstating the gauge connection yields 
	\begin{equation}\label{a}
		\begin{split}
			i\left(\partial_t+\Psi_t  +i\phi\right)\psi &= +\sigma_ze^{\Phi} m \psi  \\
			-i\sigma_ye^{\rho}&\left(\partial_x + \frac{1}{2}\Psi_x +\frac{1}{2}\Phi_x +i\partial_x \Lambda \right)\psi  \\
			-i\sigma_xe^\rho &\left(\partial_y + \frac{1}{2}\Psi_y +\frac{1}{2}\Phi_y +i\partial_y \Lambda \right)\psi.
		\end{split}
	\end{equation}
	Upon defining
	\begin{equation}\label{fieldredef}
		\chi := e^{[\frac12(\Phi + \Psi)+i\Lambda]}\psi ,
	\end{equation}
	equation (\ref{a}) reduces to 
	\begin{equation}\label{generalDirac}
		\begin{split}
			\partial_t\chi=&-ie^{\rho}\left(\sigma_y\partial_x +\sigma_x\partial_y \right)\chi  \\
			&+\left(\sigma_ze^{\Phi} m + \mathds{1}[\phi- \partial_t \Lambda] \right) \chi.
		\end{split}
	\end{equation} 
	
	We have assumed the metric to be of the form (\ref{prototype}) because this is precisely the form for which the resulting Dirac equation is implementable with our proposed scheme in a two-dimensional coupled waveguide array, below. As will be explained in section \ref{generalcase}, our scheme allows implementation of partial differential equations precisely of the form
	\begin{equation}\label{realisableeq}
		\begin{split}
			\partial_t\chi=&-ie^{\rho(x,y)}\left(\sigma_y\partial_x +\sigma_x \partial_y \right)\chi\\
			&+\left(\sigma_z \hat{h}(x,y,t)  + \mathds{1}\tilde{h}(x,y,t) \right) \chi,
		\end{split}
	\end{equation}
	in a two dimensional square photonic lattice.
	
	To see how (\ref{realisableeq}) constrains metric and vector potential, it is expedient to consider the gauge and gravity cases separately.
	For the gauge part, note that while $\sigma_z$ occurs independently of any differential operator in equation (\ref{realisableeq}), the Pauli matrices $\sigma_{x,y}$ are always multiplied by partial derivatives. Thus we see that we have to be able to gauge away the spatial  part of the vector potential. If this were not the case there would be terms in which the Pauli matrices $\sigma_{x,y}$ would appear independently of partial derivatives, since the gauge covariant derivative is given by $\nabla_\mu = \partial_\mu + i A_\mu$,
	which is contracted with the gamma-matrices $\{\gamma^\mu\}$. 
	To be able to carry out a field redefinition like (\ref{fieldredef}) that cancels the $x$ and $y$ components of the gauge potential in the differential equation, we indeed have to be able to express the spatial part as $\partial_x\Lambda dx + \partial_y \Lambda dy$ for some function $\Lambda$.
	
	For the gravity part, note that since each derivative term is multiplied by a single Pauli matrix and not a linear combination of them (at fixed position and time), the metric must necessarily be diagonal for our choice of gamma matrices. If it were not, the off-diagonal elements of the vielbein would give rise to terms such as  $\gamma^a \partial_b, \ (a\neq b)$.
	To ensure that possible terms arising from the gravity part, where $\sigma_{x,y}$ Pauli matrices would appear independently of derivatives, can be absorbed into a field redefinition such as (\ref{fieldredef}), we have to demand that the two nonvanishing entries in the spatial sector of the metric are the same.
	Note that this then also implies that we may without loss of generality consider the functions $B$ and $C$ in (\ref{realisableeq}) to be identically zero.
	Lastly, we note that after bringing the Dirac equation into its Hamiltonian form, demanding that the function $\rho= \rho(x,y)$ that modifies the strength of the derivative terms be time-independent is equivalent to assuming  $\Phi-\Psi = \rho$ to be independent of time. This is precisely the demand in (\ref{timeind}).

	\subsection{Example: Conical space and the gravitational Aharonov-Bohm effect}\label{paralleltransport}
	Before ending this section let us consider a specific example. Setting the gauge connection to zero, we assume that the metric is given by
	\begin{equation}
		\begin{split}
			ds^2& = dt^2 - r^{-2 \Delta}(dr^2+r^2 d\theta^2)\\
			&=dt^2 - \left(\sqrt{x^2+y^2}\right)^{-2 \Delta}\left(dx^2+dy^2\right).
		\end{split}
	\end{equation} 
	A short discussion on the underlying geometry is provided in Appendix  \ref{topdef}. 
	Defining
	\begin{equation}
		f:=\frac1g:= e^{-\Psi} = \left(\sqrt{x^2+y^2}\right)^{ \Delta},
	\end{equation}
	the Dirac equation in this spacetime can be written as 
	\begin{align}
		ig\partial_t \psi =& - \frac{1}{\sqrt{g}}i\sigma_y\left(\partial_x + \frac{g_x}{2g}\right)\psi -\frac{1}{\sqrt{g}}i\sigma_x\left(\partial_y + \frac{g_y}{2g}\right)\psi & \nonumber \\ &+ g \sigma_z \psi.
	\end{align}
	Upon redefining $\chi:= \sqrt{g} \psi$ this further reduces to
	\begin{equation}\label{DeString}
		i \partial_t \chi =if\left( \sigma_y \partial_x + \sigma_x \partial_y\right) \chi+ m\sigma_z \chi.
	\end{equation}
	
	Let us now explore what happens to a spinor that is parallel transported along a curve in this spacetime. Let us assume that the curve is parametrized by $\xi$, in which case the tangent vector to the curve be given by $\dot \xi$.
	To find the expression for a parallel transported spinor, we have to solve the equation
	\begin{equation}
		\nabla_{\dot \xi} \psi = \dot \xi^\mu \nabla_{\mu}\psi = 0.
	\end{equation}
	With the help of (\ref{connectionmatrices}) and by writing $\psi =V\psi_0$, where $V:=V_\xi(\tau)$ is defined to be the operator that parallel transports a spinor along the curve $\xi$, we may write this as 
	\begin{equation}
		-\frac{dV}{d \tau} = \frac{i \sigma_z}{2}\left(\dot \xi^y \Psi_x - \dot \xi^x 	 \Psi_y\right) V,
	\end{equation}
	where $\dot \xi^{x/y}$ is the $x/y$ component of $\dot \xi$. 
	Next we note that in the case at hand we have $\Psi(t,x,y) = \Psi(r)$, with $r= \sqrt{x^2+y^2}$.  Hence
	\begin{equation}
		\frac{dV}{d \tau} V^{-1} = \frac{i \partial_r \Psi \sigma_z}{2r} (\xi^x\dot \xi^y - \xi^y\dot \xi^x),
	\end{equation}
	with $r^2 = (\xi^y)^2 + (\xi^x)^2$.
	
	Assuming that we parallel transport around a concentric circle, we may write  $\xi^1= r\cos(\tau)$, $\xi^2= r\sin(\tau)$, and hence write the above equation as 
	\begin{equation}
		\frac{dV}{d \tau} V^{-1} = \frac{i r\partial_r \Psi \sigma_z}{2} (-\sin^2(\tau) -\cos^2(\tau)).
	\end{equation}
	Noting that
	\begin{equation}
		r \partial_r \Psi = -\Delta,
	\end{equation}
	one readily obtains 
	\begin{equation}
		V(\tau) = \exp\left( \Delta \frac{i\sigma_z}{2} \tau\right).
	\end{equation}
	It can in fact be shown that the assumption that we are on a concentric circle is not necessary and this expression is valid for general curves \cite{Bezerra}.
	We thus note that parallel transport of a spinor around the center once, amounts to a multiplication with
	\begin{equation}
		U(2 \pi)= \cos(\Delta \pi) + i \sigma_z\sin(\Delta\pi).
	\end{equation}

In section \ref{experiment}, we will propose an experiment to observe this non-trivial rotation in a tabletop experiment. 

\section{Effective coupling constants in Coupled waveguide arrays}\label{photlat}
We want to construct an implementation of (\ref{generalDirac}) in a coupled waveguide array. 
Our starting point for this endeavour is the usual description of the array in terms of coupled mode equations.
For a 2D square array these are given by, 
\begin{equation}
\label{cmeqn2d}
\begin{split}
i\frac{\partial \tilde c_{ab}}{\partial z} =& k_0 \left( \tilde c_{(a+1)b} + \tilde c_{(a-1)b} +\tilde c_{a(b-1)} +\tilde c_{a(b+1)} \right)\\
& + \sigma_{ab}(z) \tilde c_{ab}.
\end{split}
\end{equation}
assuming only nearest neighbour and negligible diagonal couplings  \cite{Szameit}. 
Here $z$ is the Cartesian coordinate along the waveguides,   $\tilde c_{ab}$ denotes the amplitude in the waveguide at site $(ab)$, $\sigma_{ab}(z)$ is the difference between the refractive index of the waveguide and the bulk refractive index at site $(ab)$ and position $z$.
The physical coupling strength determined by the overlap between the transverse components of the modes in adjacent waveguides is denoted by $k_0$.
Neglecting diagonal couplings is justified if the overlap of the evanescent tails between waveguides that are separated by $\sqrt2 \delta_s$ is negligible compared to the overlap for wave guides that are separated by $\delta_s$. In Appendix \ref{diagonal} we derive a scheme to effectively cancel any diagonal couplings, in a general setting. 

To allow for site-dependent coupling constants, we generalize the idea proposed in \cite{Efremidis} and introduce refractive index modulations at every site:
\begin{equation}
\sigma_{ab}(z) = \sum_{i=1}^\Lambda \frac{A_{ab}^i}{2 } \cos({\alpha_i z})
\end{equation}
Then we apply the transformation 
\begin{equation}
\tilde{c}_{ab} = c_{ab} \exp \left(i  \sum_{j=1}^\Lambda \frac{ A^j_{ab}}{2 \alpha^j}  \sin (\alpha_j z) \right),
\end{equation}
which cancels the sinusoidal terms in the coupled mode equations and obtain
\begin{align}\label{cmem}
		i\frac{\partial c_{ab}}{\partial z} =   & k^*_{[a\leftrightarrow (a+1)]b} c_{(a+1)b} 
	+	  k_{[(a-1)\leftrightarrow a]b} c_{(a-1)b} \nonumber \\
	+	&  k^*_{a[b\leftrightarrow (b+1)]}c_{a(b+1)} 
	+	  k_{a[(b-1)\leftrightarrow b]} c_{a(b-1)}.
\end{align}
Here we have defined
\begin{equation}
\begin{split}
&k_{[(a-1)\leftrightarrow a]b} =\\
& k_0  \exp \left(i\sum\limits_{j=1}^\Lambda  \frac{        A^j_{(a-1)b}-A^j_{ab}        }{2 \alpha_j}  \sin (\alpha_j z) \right)
\end{split}
\end{equation}
and 
\begin{equation}
	\begin{split}
		&k_{a[(b-1)\leftrightarrow b]} =\\
		& k_0  \exp \left(i\sum\limits_{j=1}^\Lambda  \frac{        A^j_{a(b-1)}-A^j_{ab}        }{2 \alpha_j}  \sin (\alpha_j z) \right).
	\end{split}
\end{equation}

Next we assume that the refractive index is oscillating fast. That is, the scale of variation in the $z$-direction of the modes $c_{ab}(z)$ is much larger than $1/\alpha_j$ for all $j$. 
Furthermore we choose
\begin{equation}\label{naturalspeed}
\alpha_j=q_j \omega,
\end{equation}
with $q_j$ positive natural numbers. 
Integrating (\ref{cmem}) over one modulation period $T:= \frac{2\pi}{\omega}$ under these assumptions then yields the effective coupling constant
\begin{equation}\label{effcoup}
k_{[(a-1)\leftrightarrow a]b}^{\textit{eff}} :=\frac1T\int\limits_{z_0}^{z_0 +T} k_{[(a-1)\leftrightarrow a]b}(z)dz
\end{equation}
and similarly in the $b$-direction. We now expand the integrand using the generating function of the Bessel functions of the first kind
\begin{align}\label{Bessel}
\exp \left(i \sigma \sin(x) \right) = \sum\limits_{n\in \mathds{Z}} J_n(\sigma)e^{inx},
\end{align}
for each index $j$. 
In Appendix \ref{besselapprox} we show that one may neglect contributions that contain $J_{n \neq 0}(...)$ terms.

Therefore, we may approximate our coupling constants as
\begin{equation}
\label{coupx}
k_{[(a-1)\leftrightarrow a]b} \approx k_0  \prod_{j=1}^{\Lambda} J_0\left( \frac{ A^j_{(a-1)b}-A^j_{ab}}{2 \alpha^j}\right)
\end{equation}
and
\begin{equation}
\label{coupy}
k_{a[(b-1)\leftrightarrow b]} \approx k_0  \prod_{j=1}^{\Lambda} J_0\left( \frac{ A^j_{a(b-1)}-A^j_{ab}}{2 \alpha^j}\right).
\end{equation}
Later, we will need to make the coupling between neighbouring waveguides position dependent. Suppose we are given a scalar function $h$ on the lattice sites:
\begin{equation}
0 < h(a,b) \leq \max_{c,d} h(c,d) 
\end{equation}
For reasons that will become apparent soon, we would like to choose the coupling between sites to be proportional the geometric mean of the values given by the values the function $h$ takes at the respective sites times the original coupling constant. This would for example require the coupling between $c_{ab}$ and $c_{(a+1)b}$ to equal $k \frac{\sqrt{h(a,b)h((a+1),b)}}{ \max_{c,d} h(c,d) }$.
To achieve this we introduce two refractive index modulations, whose amplitudes we will denote by $A_{ab}$ and $B_{ab}$ respectively and whose modulation speeds are denoted by $\alpha$ and $\beta$.
For the "$A$" modulations we assign amplitudes according to
\begin{equation}\label{Aassignment}
\left[\begin{matrix}
0 & \ & A_{(a+1)(b+1)}   \\
\ & \ & \         \\
A_{ab} & \ & 0   \\
\end{matrix}\right],
\end{equation}
extended periodically over the entire lattice and where $A_{ab}$ is chosen such that
\begin{equation}
J_0\left(\frac{A_{ab}}{2\alpha}\right)=\sqrt{\frac{h(a,b)}{\max_{c,d} h(c,d)}}.
\end{equation}
For the "$B$" modulation, we assign amplitudes according to 
\begin{equation}\label{Bassignment}
\left[\begin{matrix}
B_{a(b+1)}  & \ & 0  \\
\ & \ & \         \\
0 & \ & B_{(a+1)b}  \\
\end{matrix}\right],
\end{equation}
with 
\begin{equation}
J_0\left(\frac{B_{ab}}{2\beta}\right)=\sqrt{\frac{h(a,b)}{\max_{c,d} h(c,d)}}.
\end{equation}
This then yields coupling constants according to 
\begin{equation}
k_{[a \leftrightarrow (a+1)]b} =k   \frac{\sqrt{h(a,b)h((a+1),b)}}{ \max_{c,d} h(c,d) }
\end{equation}
and similarly in the $b$-direction.

\section{Waveguide implementation of The Dirac equation}\label{impl}
In this section we describe how one may implement the Dirac equation in a coupled waveguide array. Let us first consider the case without gravity and electromagnetism.

\subsection{The flat, uncharged case}\label{freediracequation}

The Dirac equation for a free particle in 2+1 dimensional flat spacetime can be written as
\begin{equation}\label{flatDElat}
i \partial_t \Psi = -i \left(\sigma_y \partial_x + \sigma_x \partial_y \right)\Psi + m \sigma_z \Psi.
\end{equation}
To bring this equation onto a square lattice, we will employ equations (\ref{coupx}) and (\ref{coupy}).
We denote the discretized two-spinor components as
\begin{equation}
\prescript{\ }{j}{\psi}_{n}^{m}:=\Psi_j\left(\delta_x + n\delta_x ,\delta_y+m\delta_y\right).
\end{equation}
On a `unit cell', let us try assigning spinor components as 
\begin{equation}\label{initialprescription}
\left[\begin{matrix}
\prescript{\ }{1}{\psi}_{n-1}^{m+2} & \prescript{\ }{2}{\psi}_{n}^{m+2} & \prescript{\ }{1}{\psi}_{n}^{m+2} &  \prescript{\ }{2}{\psi}_{n+1}^{m+2}\\
i\prescript{\ }{2}{\psi}_{n-1}^{m+1} & i\prescript{\ }{1}{\psi}_{n}^{m+1} & i\prescript{\ }{2}{\psi}_{n}^{m+1}& i\prescript{\ }{1}{\psi}_{n+1}^{m+1} \\
\prescript{\ }{1}{\psi}_{n-1}^{m} & \prescript{\ }{2}{\psi}_{n}^{m} & \prescript{\ }{1}{\psi}_{n}^{m} & \prescript{\ }{2}{\psi}_{n+1}^{m}  \\
i\prescript{\ }{2}{\psi}_{n-1}^{m-1} & i\prescript{\ }{1}{\psi}_{n}^{m-1} & i\prescript{\ }{2}{\psi}_{n}^{m-1} & i\prescript{\ }{1}{\psi}_{n+1}^{m-1}  \\
\end{matrix}\right]
\end{equation}
which is to be periodically extended over the entire lattice.
The derivatives are discretized as
\begin{equation}
\partial_y \Psi_i \approx  \frac{\Psi_i\left(n \delta_x ,(m+1) \delta_y\right) - \Psi_i(n\delta_x ,(m-1)\delta_y)}{2\delta_y} 
\end{equation}
and
\begin{equation}
\partial_x \Psi_i \approx  \frac{\Psi_i\left((n+1) \delta_x ,m \delta_y\right) - \Psi_i\left(n\delta_x ,m\delta_y\right)}{\delta_x}
\end{equation}
or
\begin{equation}
\partial_x \Psi_i \approx  \frac{\Psi_i(n \delta_x ,m \delta_y) - \Psi_i((n-1)\delta_x ,m\delta_y)}{\delta_x},
\end{equation}
depending on the spinor component "$i$" and the column of the lattice.

Now suppose that the coupling constants between any two given lattice sites always have the same absolute value. Then, to realize the discretized version of (\ref{flatDElat}), we necessarily have to set the absolute value of our coupling constant to
\begin{equation}
|k| = \frac1{\delta_x}=\frac1{2\delta_y}.
\end{equation}
A quick calculation shows that the signs of the coupling constants have to vary as
\begin{equation}
\label{couplingsign}
\left[\begin{matrix}
\psi & - & \psi & +&\psi &- &\psi    \\
   + & \ & +    &\ & +   &\ & +         \\
\psi & + & \psi & -&\psi &+ &\psi    \\
-    & \ & -    &\ & -   &\ &  -     \\
\psi & - & \psi & +&\psi &- &\psi   \\
+    & \ & +    &\ & +   &\ & +      \\
\psi & + & \psi & -&\psi &+ &\psi  \\
\end{matrix}\right].
\end{equation}
As it turns out, it is impossible to achieve such a prescription using only a single modulation of the refractive index. Since our effective coupling constants scales with the number of introduced modulations $\Lambda$ as   $\left[J_0(...)\right]^\Lambda$, we want to minimize the number of introduced refractive index modulations.

To achieve this goal we note that multiplying an entry in (\ref{initialprescription}) with $(-1)$ changes the sign of the immediately surrounding coupling constants that need to be achieved. This means that if we change (\ref{initialprescription})  to
\begin{equation}\label{prescription}
\left[\begin{matrix}
-\prescript{\ }{1}{\psi}_{n-1}^{m+2} & \prescript{\ }{2}{\psi}_{n}^{m+2} & \prescript{\ }{1}{\psi}_{n}^{m+2} &  -\prescript{\ }{2}{\psi}_{n+1}^{m+2}\\

i\prescript{\ }{2}{\psi}_{n-1}^{m+1} & -i\prescript{\ }{1}{\psi}_{n}^{m+1} & -i\prescript{\ }{2}{\psi}_{n}^{m+1}& i\prescript{\ }{1}{\psi}_{n+1}^{m+1} \\

\prescript{\ }{1}{\psi}_{n-1}^{m} & -\prescript{\ }{2}{\psi}_{n}^{m} & -\prescript{\ }{1}{\psi}_{n}^{m} & \prescript{\ }{2}{\psi}_{n+1}^{m}  \\
-i\prescript{\ }{2}{\psi}_{n-1}^{m-1} & i\prescript{\ }{1}{\psi}_{n}^{m-1} & i\prescript{\ }{2}{\psi}_{n}^{m-1} & -i\prescript{\ }{1}{\psi}_{n+1}^{m-1}  \\
\end{matrix}\right],
\end{equation}
(\ref{couplingsign}) changes to
\begin{equation}
\left[\begin{matrix}
\psi & + & \psi & +&\psi &+ &\psi   \\
- & \ & -   &\ & -  &\ & -      \\
\psi & - & \psi & -&\psi &- &\psi   \\
-    & \ & -    &\ & -   &\ &  -        \\
\psi & + & \psi & +&\psi &+ &\psi   \\
-   & \ & -    &\ & -   &\ & -       \\
\psi & - & \psi & -&\psi &- &\psi   \\
\end{matrix}\right],
\end{equation} 
which can be achieved with only one sinusoidal modulation
\begin{equation}
\sigma_{mn}(z)=\frac{C_{mn}}{2}\cos(\gamma z).
\end{equation}
To specify the required amplitude prescription, we note that the relation
\begin{equation}
J_0\left(\xi\right) = -J_0\left(3\xi\right)
\end{equation}
has a non-empty set of solutions. The first few positive roots are given by $\xi_1 \approx  2.704$, $\xi_2 \approx  5.83$ and $\xi_3 \approx 8.97$ and e.g.~$J_0(\xi_1)=-J_0\left(3\xi_1 \right) \approx -0.144.$
We define  $C:= \xi_1/2\gamma$ and assign the amplitudes as
\begin{equation}\label{signcoupling}
\left[\begin{matrix}
	0 & + & 3C & +&0 &+ &3C   \\
	- & \ & -   &\ & -  &\ & -        \\
	C & - & 2C & -&C &- &2C  &  \\
	-    & \ & -    &\ & - & \  &-         \\
		0 & + & 3C & +&0 &+ &3C   \\
	-   & \ & -    &\ & - &\  &-        \\
	C & - & 2C & -&C &- &2C   \\
\end{matrix}\right].
\end{equation} 
where the signs of the coupling constants follow from the assigned amplitudes by equations (\ref{coupx}) and (\ref{coupy}).

We have thus realized the discretized massless $2 + 1$ dimensional Dirac equation with
\begin{equation}
J_0\left(3\xi_1\right)=|k| = \frac1{\delta_x}=\frac1{2\delta_y}.
\end{equation}
If we would like to take $ \delta_x =  \delta_y$, so that the discretisation lengths in the two directions coincide, it is obvious from the above considerations that this corresponds to introducing different coupling constants in the $x$ and $y$ directions:
\begin{equation}
k_x=\frac12 k_y. 
\end{equation}
This can be achieved by changing the distance of the waveguides in an experimental setup, or by introducing additional refractive index modulation in the $x$ direction similarly for every $y$ column, according to $[0,E]$ extended periodically so that $J_0\left(\frac{E}{2 \epsilon}\right) = \frac12$. This changes the effective coupling in the $x$ direction to one half of its original size, while leaving the coupling in the $y$ direction invariant.
We will however not do so and instead work with different discretisation lengths in the $x$ and $y$ directions, so that the coupling constants in both directions can have the same value.
Including a mass term can be achieved through the usual method of alternating high and shallow refractive indices \cite{Dreisow10,Dreisow12}.

\subsection{The general case}\label{generalcase}

Our starting point is (\ref{generalDirac}), which we repeat here for convenience:
\begin{equation*}
\begin{split}
\partial_t\chi=&-ie^{\rho}\left(\sigma_y\partial_x +\sigma_x\partial_y \right)\chi \\
&+\left(\sigma_ze^{\Phi} m + \mathds{1}[\phi- \partial_t \Lambda] \right) \chi
\end{split}
\end{equation*}
Just like the mass term, $\left(\sigma_ze^{\Phi} m + \mathds{1}[\phi- \partial_t \Lambda] \right)$ can be easily implemented by changing the refractive index of the respective waveguide appropriately.
Hence we turn to the derivative term.
For ease in notation we define $f:= e^\rho$. 
The relevant part of the equation may thus be written as
\begin{equation}\label{releq}
i\partial_t\chi= -if\left(\sigma_y  \partial_x + \sigma_x \partial_y  \right) \chi
\end{equation}
If we want to put this on the lattice, we will have to make the absolute values of the effective coupling constants position dependent. Furthermore, we need to be careful to obtain the correct differential equation in the limit of vanishing discretization length.
For this, the effective couplings of the spinor components (as opposed to the amplitudes in waveguides) to its neighbours are the determining factors.

It is easy to verify that in the discretized version of (\ref{releq}) the coupling of a spinor component $\prescript{\ }{}{\psi}_{n}^{m}$ to its neighbours $\prescript{\ }{j}{\psi}_{k}^{l}$ must be local, i.e.~the absolute value may only depend on the tuple $(nm)$ and must not depend on the coordinates of the neighbouring sites $(kl)$. 

To achieve this, we will use the lattice constructed in the previous subsection, superimposed with the modulation of coupling constants that we introduced at the end of section \ref{photlat}.
Note that while we were considering a function $h(a,b)$ on the lattice sites $\{(ab)\}$ there, we are now considering a function $f(n,m)$ on discretized space points $\{(nm)\}$. Given the prescription (\ref{prescription}) of spinor components, the two are related as
\begin{equation}
h(a,b) := f(\lfloor  a/2 \rfloor,b),
\end{equation}
as is easily seen.
Here $\lfloor\cdot\rfloor$ denotes the floor function that assigns to each real number its closest smaller integer.

Additionally we modify the assignment of spinor components to our lattice by  pointwise multiplication with 
\begin{multline}\label{Zfactor0}
 \left[\begin{matrix}
Z_{(n-1)(m+2)} & Z_{n(m+2)}& Z_{n(m+2)} &  Z_{(n+1)(m+2)}&     \\
Z_{(n-1)(m+1)} & Z_{n(m+1)}& Z_{n(m+1)} & Z_{(n+1)(m+1)} &\\
Z_{(n-1)m} & Z_{nm}& Z_{nm} & Z_{(n+1)m}& \\
Z_{(n-1)(m-1)} & Z_{n(m-1)}&  Z_{n(m-1)}  & Z_{(n+1)(m-1)}&\\
\end{matrix}  \right]
\end{multline}
extended periodically.
We now define
\begin{equation}\label{Zfactora}
\sqrt{\frac{f(n,m)}{f(n+1,m)}} Z_{nm} = Z_{(n+1)m}
\end{equation}
and likewise
\begin{equation}\label{Zfactorb}
\sqrt{\frac{f(nm)}{f(n,m+1)}} Z_{nm} = Z_{n(m+1)}.
\end{equation}
The consistency of these conditions is easily verified.
We have one degree of freedom left by choosing the initial $Z$ value in the inductive definition above.
Let us fix 
\begin{equation}
Z_{00}=\frac{1}{\sqrt{f(0,0)}}.
\end{equation}
Then we have
\begin{equation}\label{Zvalue}
\begin{split}
Z_{mn}&=\frac{1}{\sqrt{f(0,0)}}\cdot \prod\limits_{i=1}^n  \sqrt{\frac{f(i-1,0)}{f(i,0)}} \cdot \prod\limits_{j=1}^n \sqrt{\frac{f(n,j-1)}{f(n,j)}}\\
&=\frac{1}{\sqrt{f(n,m)}}.
\end{split}
\end{equation}
Note that we could have chosen any other path from $(0,0)$ to $(nm)$.
Plugging in the relations  (\ref{Zfactor0}) -  (\ref{Zfactorb}) into the effective coupled mode equations that we constructed in Section \ref{freediracequation} then shows that we have indeed realized the discretized version of equation (\ref{releq}) if we choose the initial coupling constant $k_0$ according to 
\begin{equation} 
k_0:=  \frac{\max_{m,n} f(n,m)}{\delta_x} = \frac{\max_{m,n} f(n,m)}{2\delta_y}.
\end{equation}

Checking that the discretisation scheme yields the correct differential equation in the limit of vanishing stepsize is straightforward, since the function multiplying the discretized derivative term is local.
Lastly we note that our implementation scheme is able to precisely realize equations of type (\ref{realisableeq}). The only part that is not immediately obvious is 
the relationship between the functions multiplying the spatial derivative terms. 
To unravel this relationship, assume that the $x$-derivative is multiplied by $f=e^{q(x,y)}$ and the $y$ derivative is multiplied by $g=e^{p(x,y)}$.
For consistency we must then require
\begin{equation}
\frac{g_{n,m}}{g_{n,m+1}} \frac{f_{n,m+1}}{f_{n+1,m+1}}=\frac{f_{n,m}}{f_{n+1,m}}\frac{g_{n+1,m}}{g_{n+1,m+1}}
\end{equation}
for arbitrary $\delta_x,\delta_y$.
Taking the logarithm of this equation, dividing by $ \left(\delta_x  \delta_y\right)$ and taking the limits $\delta_x,\delta_y \rightarrow 0$ then implies
\begin{equation}
\partial_x \partial_y q = \partial_x \partial_y p
\end{equation}
by Taylor's theorem.
Hence
\begin{equation}
p - q =  B(x) + C(y) ,
\end{equation}
for some functions $B$ and $C$ of only a single variable.
Thus the partial differential equations that are implementable indeed take the form of (\ref{realisableeq}).

Now that we have shown, in principle, how the Dirac equation in classical backgrounds may be implemented in waveguide arrays, we would like to draw the reader's attention to the considerable engineering challenge associated with actually fabricating the waveguide arrays needed for our implementation scheme:
In addition to the non trivial coupling between the spinor components, we have encoded the entire information about the mass of our test particle, the background spacetime and the electromagnetic background it experiences into the refractive indices of the waveguides. The limiting case of a flat spacetime and the gauge field set to zero is certainly realisable, as there is only a single sinusoidal refractive index modulation required (for an experimental realisation of a similar setup see e.g. \cite{Szameit09}).
It has to be investigated on a case by case basis for any given more complex background scenario if it might be realised with state of the art techniques.

\section{Gravitational Aharonov-Bohm effect}\label{experiment}

In this section, we devise a thought experiment to observe the gravitational Aharonov-Bohm effect. We then show that an analogue can be implemented in a tabletop experiment with the methods constructed above. 
\subsection{Theoretical considerations}\label{extheo}
Our starting point is the conical space. As argued in Appendix \ref{topdef}, we may write the metric as
\begin{align}
ds^2 &= dt^2 - \left(\sqrt{x^2+y^2}\right)^{-2 \Delta}\left(dx^2+dy^2\right),
\end{align}
in a global coordinate system.
Going to polar coordinates, we can write this as
\begin{equation}
ds^2 = dt^2-R^{-2\Delta}\left(dR^2+R^2 d \tau^2\right).
\end{equation}
The spacetime is locally flat, i.e. every point has an open coordinate neighbourhood on which the metric takes Minkowskian form. While there is no such coordinate chart that covers all of the cone, one may however introduce coordinate neighbourhoods in which the metric is Minkowskian, that cover the entire cone except for a straight line starting at the tip of the cone. For fixed time, this can be understood geometrically as cutting the cone along this line and flattening out the obtained surface \cite{Vilenkin}. 

The result is a copy of $\mathds{R}^2$ with a wedge of angular size $\Delta2\pi$ removed. In such a coordinate system, the metric is given by 
\begin{equation}
ds^2 = dt^2-d\rho^2 - \rho^2 d\theta^2
\end{equation}
with
\begin{equation}
0 < \theta < (1-\Delta) 2 \pi.
\end{equation}
The points for which $\theta = 0$ and $\theta = (1-\Delta) 2 \pi$, i.e.~at the edge of the removed wedge, would correspond to the line along which we cut the cone and need to be excluded. This is for the same reason one excludes the values $\{0, 2\pi\}$ in the usual polar coordinates of $\mathds{R}^2$. To be mathematically precise, this ensures that our coordinate maps are  diffeomorphisms between open sets. Thus we see that to describe the entire cone using only flat coordinate systems, we need an additional flat coordinate patch. This corresponds to cutting the cone along a different line. To signify the difference between these coordinate systems, we denote the angular variable by $\phi$ for the second coordinate system. As before, we have
\begin{equation}
ds^2 = dt^2-d\rho^2 - \rho^2 d\phi^2
\end{equation}
with
\begin{equation}
0 < \phi < (1-\Delta) 2 \pi.
\end{equation}

\begin{figure}[ht!]
	\begin{center}
		\includegraphics[width=0.4\textwidth]{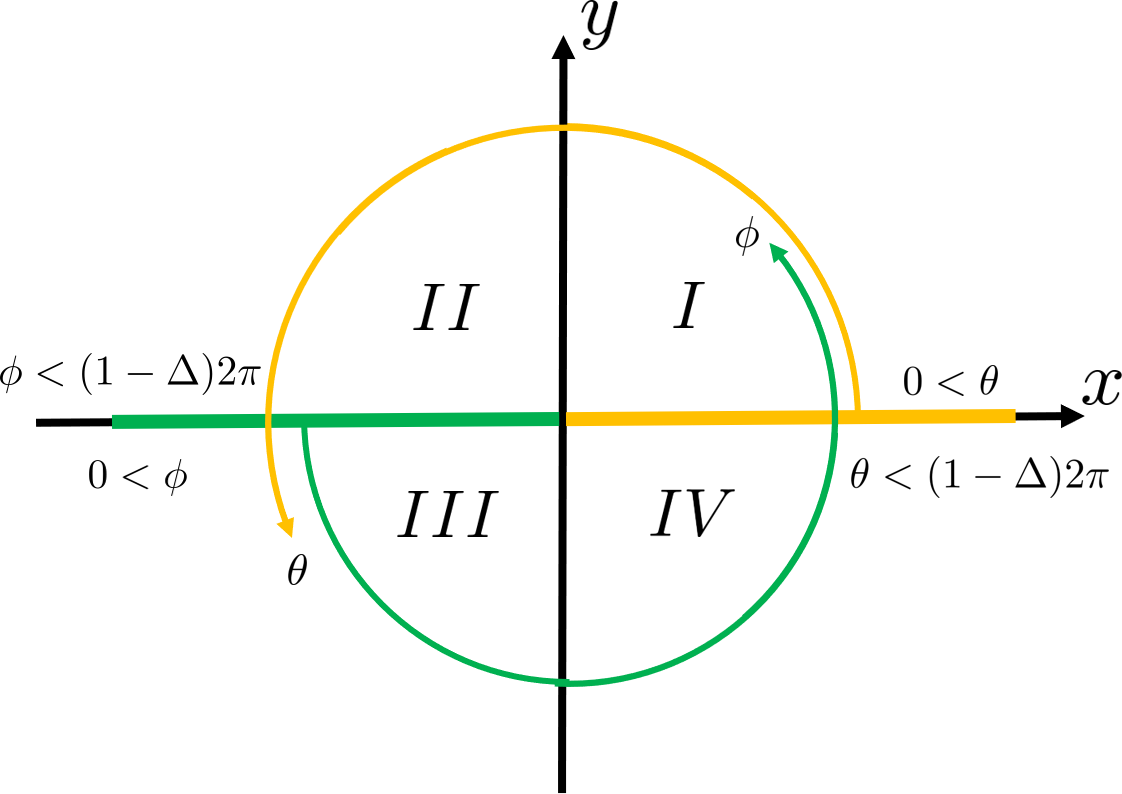}
	\end{center}
	\caption{Coordinate systems. The colored sections of the $x$-axis mark the half-line on the cone which is not covered by the flat coordinate systems.}
	\label{coords}
\end{figure}

We make the cuts along the positive and negative $x$ axes of our global coordinate system (see Figure \ref{coords}). 
The idea now is to mimic the Aharonov-Bohm effect for $U(1)$ gauge fields. For this we want to create an input state that corresponds to two  separately localised parts $\Psi_0, \Phi_0$  of a wavefunction with initial momenta chosen such that one of the localised parts will travel above the tip of the cone at $(x=0,y=0)$, while the other travels below. Heuristically, we think of the two areas of localisation of the wavefunction as two separate particles

\begin{figure}[ht!]
	\begin{center}
		\includegraphics[width=0.4\textwidth]{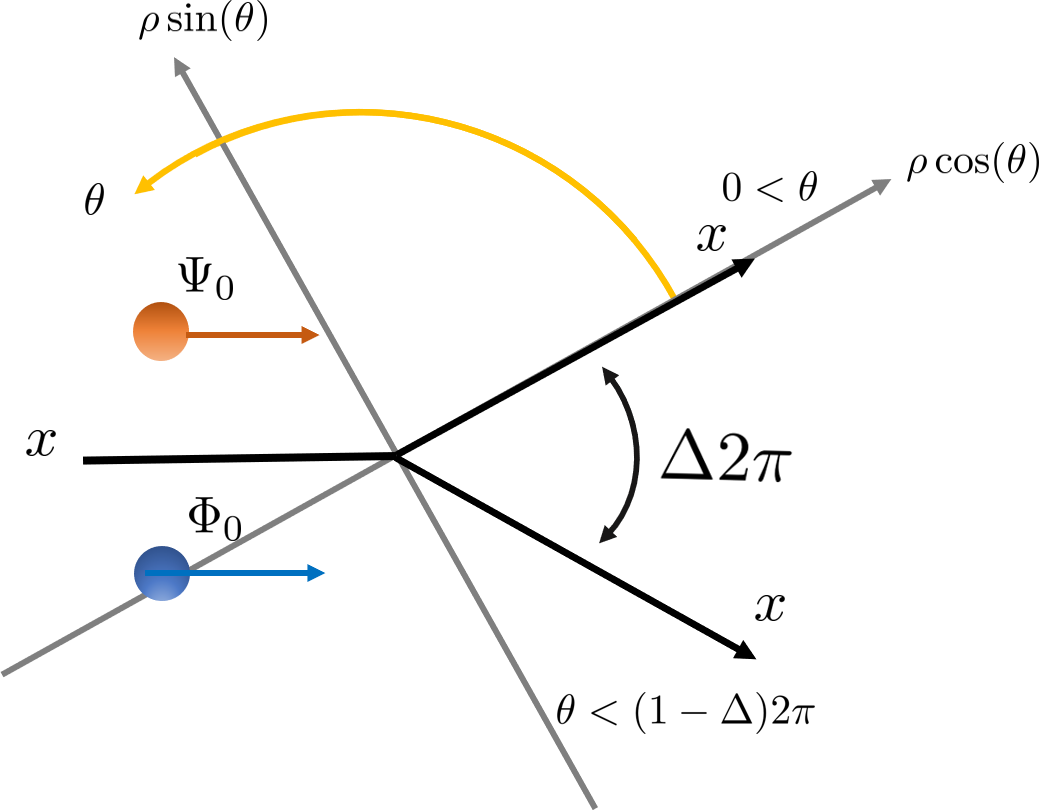}
	\end{center}
	\caption{Input state preparation in a flat coordinate system. The input states $\Psi_0,\Phi_0$ are taken to be localised and centered around momenta  parallel to the half-line that corresponds to $x<0$ in the global coordinate system, signified by the arrows. Note that a half line starting at zero in the global coordinate system still corresponds to a half line in the flat coordinate system. }
	\label{initialstate}
\end{figure}

In flat spacetime it is well known how to create  localised  states with  Gaussian momentum distribution. One e.g.~projects a plane wave solution in momentum space onto the positive energy subspace, multiplies with a Gaussian centered around the desired momentum and Fourier transforms \cite{Thaller}.
We understand the free time evolution of such initial states in the flat spacetime and can easily calculate it numerically. We can use this knowledge to understand the dynamics in conical space, by making use of the fact that our spacetime is locally flat.
We initially work in the coordinate system characterised by the angular variable $\theta$ and prepare an input state
\begin{equation}
\Psi_\text{flat} := \Psi_0 + \Phi_0
\end{equation}
normalised as $||\Psi_0||=||\Phi_0||=\frac1{\sqrt 2}$.
 $\Psi_0$ is taken to be localised in sector $II$ and $\Phi_0$ in sector $III$ (compare figure \ref{initialstate}).

We choose the $\theta$-coordinate system for the preparation of the input states, since we heuristically think about the two localised parts of the initial state as originating from a common source located in the direction opposite to the initial momenta of the two parts. Hence we need a coordinate patch that covers the entire geometry of such a setup, including the point of the source, the parts of the spacetime where the states $\Psi_0,\Phi_0$ are localised and the trajectory between the source and the state. We want to describe the subsequent time evolution in the coordinate system characterised by the angular variable $\phi$, since we will mainly be concerned with observations in the sectors $I$ and $IV$ and hence want a flat coordinate system that is able to describe this patch. 

To be able to do so, we require the time evolved states to not extend through the line along which we cut our cone to go to the $\phi$-coordinate system (i.e.~$\{x<0, y=0\}$ under the identification of the constant time slice of the manifold with the global coordinate system). 
Fortunately enough this can be guaranteed:
Let us first approximate the initial Gaussian wavefunction by zero whenever the values fall below a certain threshold. This can be done smoothly, if we subsequently mollify around the cutoff. Since the time evolution in flat space is implemented as a unitary operator, the time evolved truncated state approximates the time evolved true state with the initial error in norm at all times on the flat patch. Furthermore we note that the Dirac equation imposes a finite propagation speed. That is, for continuously differentiable $\psi \in C^1(\mathds{R}^2)$ which is non-zero only within a ball of radius $r$ centered at $x_0$ (i.e.~$\text{supp} \psi(0,\cdot) \subset B_r(x_0)$), we have $\text{supp} \psi(t,\cdot) \subset B_{r+t} (x_0)$, i.e.~it is non-zero only within a Ball of radius $(r+t)$ \cite{Lienert}. 
Hence if we prepare the initial states around large enough initial momenta and far enough away from the half line $\{x<0, y=0\}$, we avoid the truncated states to reach this half line  before the expectation values have propagated beyond the tip of the cone in this coordinate system.

Hence we may change our coordinate system to the one characterized by the angular variable $\phi$ and do the subsequent calculations in this coordinate system.
We have to first check how the input states change under such a coordinate transformation to find their form in the new coordinates. 
For this, we need to know the angular change $\phi - \theta$.
We obtain this information by making reference to the angular variable $\tau$ of the global coordinate system.
From the relation of metrics in the respective coordinate systems (cf. Appendix \ref{topdef}), we can infer that where the coordinate systems overlap
\begin{equation}\label{angulardiffe}
d\phi = (1-\Delta) d \tau = d \theta .
\end{equation}
Furthermore the half line
\begin{equation}
\left\{\lim\limits_{\tau\downarrow 0^+}R(\cos(\tau),\sin(\tau))^T\  \bigg|\ R>0 \right\}
\end{equation}
in the coordinate system characterized by $\tau$ corresponds to   
\begin{equation}
\left\{\lim\limits_{\theta\downarrow 0^+}\rho(\cos(\theta)\sin(\theta))^T \  \bigg|\  \rho > 0 \right \}
\end{equation}
in the flat coordinate system characterized by $\theta$ (see Figure \ref{coords}). Here
\begin{equation}
\rho = (1-\Delta)^{-1}R^{1-\Delta}.
\end{equation}

Strictly speaking we need to take the limits on the manifold itself and not in the coordinate representation, since as argued above, the limits of the considered sequences are not in the coordinate images anymore. We however stick to this notation since it is less cluttered and there is no room for confusion.

For $y<0$ we know that the half line 
\begin{equation}
\left\{R(\cos(\pi),\sin(\pi))^T\  \bigg|\ R>0 \right\}
\end{equation}
in the coordinate system characterized by $\tau$ corresponds to the half line
\begin{equation}
\left\{\lim\limits_{\phi\downarrow 0^+}\rho(\cos(\phi),\sin(\phi))^T \  \bigg|\  \rho > 0 \right \}.
\end{equation}
Integrating the differential equations (\ref{angulardiffe}) with these initial conditions then yields
\begin{align}
\theta &= (1-\Delta)\tau, \\
\phi&=(1-\Delta)(\tau-\pi)\label{rela}.
\end{align}
For the set where $y>0$, we know that 
\begin{equation}
\left\{R(\cos(\pi),\sin(\pi))^T\  \bigg|\ R>0 \right\}
\end{equation}
corresponds to 
\begin{equation}
\left\{\lim\limits_{\phi\uparrow (1-\Delta)2\pi}\rho(\cos(\phi)\sin(\phi))^T \  \bigg|\  \rho > 0 \right \}.
\end{equation}
Together with (\ref{angulardiffe}), this then implies
\begin{align}
\theta&=(1-\Delta)\tau\\
\phi&=(1 - \Delta)(\tau +  \pi)\label{relb}.
\end{align}

An easy calculation then shows
\begin{equation}\label{trafo1}
\left[\phi - \theta\right]_{\text{mod}2\pi} = (1-\Delta)\pi =: \alpha
\end{equation}
for $y> 0$, and
\begin{equation}\label{trafo2}
\left[\phi - \theta\right]_{\text{mod}2\pi} =(1+\Delta)\pi =: \beta
\end{equation}
for $y< 0$.
Of course we can infer the value of the change in the angular variable only up to $2 \pi$. 
Since we are working with spinors, this is not a triviality. As we will see momentarily, it does not pose an obstacle though.

Under a Lorentz transformation $\Lambda$, infinitesimally characterized by $\omega_{ab}(\Lambda)$, spinors change according to \cite{Peskin}
\begin{equation}
\psi(x) \longmapsto \Lambda_\frac12(\omega_{ab}(\Lambda))\psi(\Lambda^{-1}x).
\end{equation}
Here, we are working with the spin-$\frac12$ representation, denoted by $\Lambda_\frac12(\cdot)$.
Given the Clifford algebra structure of $\{\gamma_0,\gamma_1,\gamma_2\}$, we can represent the effect of a Lorentz transformation $\Lambda$ through \cite{Peskin}
\begin{equation}\label{rep}
\Lambda_\frac12(\omega_{ab}(\Lambda) ) := \exp\left( \frac{i}{2} \omega_{ab}(\Lambda)\frac{i}{4}[\gamma^a,\gamma^b]\right).
\end{equation}
In our case of a rotation $\Lambda = R_2(\eta)$ in the plane, the only non-vanishing entries of $\omega_{ab}$ are $\omega_{12}=-\omega_{21}= \eta$. This then yields 
\begin{equation}\label{change}
V(\eta) :=\Lambda_\frac12(\omega_{ab}(\eta)) = \exp\left( \frac{i}{2} \eta \sigma_z\right).
\end{equation}
The two coordinate systems dissect the underlying spatial slice of our manifold into two open disjoint sub-spaces, the one where $y >0$ and the one where $y<0$. Thus we a priori do not know to which of the two possible elements $\{V(\eta),V(\eta + 2 \pi)) \}$ the elements  $ R_2(\eta) \in SO(2)$, where $\eta = \alpha,  \beta$, need to be lifted for a consistent prescription. We can however infer this by considering the case $\Delta \rightarrow 0$ for which the transformation for $\eta=\alpha$ needs to coincide with the transformation for $\eta = \beta$. As this is the case for the choice of representations of the angular variables  in (\ref{trafo1}) and (\ref{trafo2}) we see that we are consistent in simply inserting $\eta=\alpha, \beta$ into (\ref{change}).
We therefore infer that our input states must change according to
\begin{equation}
\begin{split}
&\Psi_0(x) \longmapsto  V((1-\Delta)\pi)\Psi_0\left(R_2^{-1}((1-\Delta)\pi)x\right)\\
&\Phi_0(x) \longmapsto V((1+\Delta)\pi) \Phi_0\left(R_2^{-1}((1+\Delta)\pi)x\right).
\end{split}
\end{equation}

Next, let us see how to describe the time evolution of the input states that we have in the coordinate system with angular variable $\phi$.
As was to be expected, we find that it does not matter if one first time evolves and then changes the coordinate system, or if one first changes the coordinate system and then time evolves in the new coordinate system.
To see this rigorously, let us consider a Lorentz transformation $\Lambda: \mathds{R}^{1,2} \rightarrow \mathds{R}^{1,2}$, mapping from the coordinates $\{y'^\mu\}$ to $\{z^\mu\}$. Recalling the covariance of the flat Dirac equation, we know that
\begin{equation}\begin{split}
\left(i \gamma^\mu(\Lambda^{-1})\indices{^\nu_\mu} \frac{\partial}{\partial z^\nu}-m\right)\Lambda_\frac12(\omega_{ab}(\Lambda)) \psi(t,\Lambda^{-1}z)\\
= \Lambda_\frac12(\omega_{ab}(\Lambda)) \left(i \gamma^\mu \frac{\partial}{\partial y'^\mu}-m\right) \psi(t,y') =0
\end{split}
\end{equation}
holds for solutions $\psi(t,x)$ \cite{Peskin}. In our case, $t=z^0=y'^0$ and the transformation $\Lambda$ is a pure rotation $R_2(\eta):\mathds{R}^2 \rightarrow \mathds{R}^2$.
Since in this case $[ \gamma_0 , \Lambda_\frac12(\omega_{ab}(\eta))]=0$, which in our choice of representation is easily seen since $\gamma_0 = \sigma_z$ and $\Lambda_\frac12(\omega_{ab}(R_2(\eta))) = e^{\frac{i}{2}\eta\sigma_z}$, we have for the time evolution $U_z(t)$ in the $z$-coordinate system
\begin{equation}
U_z(t)V(\eta) \Psi_0\left(R_2^{-1}(\eta)z\right)
= V(\eta) \Psi(t,R_2(-\eta)z).
\end{equation}
where $\Psi(t,y') = U_{y'}(t) \Psi_0(y')$ is the solution of the Dirac equation in the $y'$-coordinate system. 

Now first take two copies of $\mathds{R}^2$ with coordinates
\begin{align}
(y'^1,y'^2) &:= r(\cos(\theta),\sin(\theta)) \ :\ 0< \theta < 2 \pi  \\
(z^1,z^2) &:= r(\cos(\phi),\sin(\phi)) \ : \ 0< \phi < 2 \pi,
\end{align}
Then we can think of the coordinate representations that describe parts of our manifold (i.e.~the ones where $0< \phi,\theta < (1-\Delta) 2 \pi)$) as the open subsets of these copies of $\mathds{R}^2$ in which the corresponding angular wedge is removed.
If we take $\eta = \alpha$, we can completely describe the time evolution of the initial $\Psi_0$ in this setting:
\begin{equation}
\begin{split}
U_z(t)V(\alpha) \Psi_0(R_2^{-1}(\alpha)z)
= V(\alpha) \Psi(t,R_2(-\alpha)z).
\end{split}
\end{equation}
We only have to make sure that at no point, after transforming the state to the $z$-coordinate system are there non-zero contributions in the wedge $(1-\Delta)2 \pi \leq \phi \leq 2 \pi$ in order to maintain a consistent description.
We argue similarly for the case of the initial $\Phi_0$; only here we have to set $\eta=\beta$:
\begin{equation}
\begin{split}
U_z(t)V(\beta) \Phi_0(R^{-1}(\beta)z)
= V(\beta) \Phi(t,R_2(-\beta)z).
\end{split}
\end{equation}

In total, assuming that neither state is non-zero in the missing wedge in the $z$-coordinate system, we have as our total state at time $t$
\begin{align}
\Psi_{\text{tot}}(t,z) = V(\alpha) \Psi(t,R_2(-\alpha)z) +V(\beta) \Phi(t,R_2(-\beta)z).
\end{align}
The probability density everywhere except on the half line $ \{ y=0,x<0 \}$ and thus especially in the quadrants $I$ and $IV$, expressed in coordinates $z^1=r\cos(\phi)$, $z^2=r\sin(\phi)$ is then given by
\begin{equation}\label{probdens}
\begin{split}
&\rho_{\Psi+\Phi}(t,z)= \bar \Psi\Psi(t,R_2(-\alpha)z)+\bar \Phi \Phi(t,R_2(-\beta)z)\\
& + 2\Re\{\bar\Psi(t,R_2(-\alpha)z) e^{i\sigma_z\Delta \pi}\Phi(t,R_2(-\beta)z)  \}.
\end{split}
\end{equation}
This dependence on the deficit angle is precisely what we expected after our discussion in section \ref{paralleltransport}. 

We note the similarity to the corresponding result for the interference pattern of electrons under the influence of the electromagnetic Aharonov-Bohm effect. There, the role of the deficit angle is played by the magnetic flux:
For the Aharonov-Bohm effect, one can obtain time-dependent solutions by time evolving initial states as if no gauge potential is present as long as we don't have to consider paths encircling the solenoid containing the magnetic field (i.e.~in any region of the space with vanishing fundamental group). Then, to describe the interference pattern of parts of the wavefunction that have `travelled' above and below the solenoid, we need to gauge transform to a common gauge choice. Since the gauge fields can't be set to zero globally, this  modifies the time evolved states by adding a phase corresponding to whether the states travelled above or below the solenoid \cite{Nakahara}.

In the gravitational case, we time evolve initial states as if our flat coordinate system extends indefinitely. This description is valid as long as we don't have to consider paths that encircle the tip of the cone. 
To calculate the probability density on the side of the cone opposite to where we prepared the initial state, we go to the coordinate system that completely covers this side of the cone.

Spinors transform non-trivially under rotations; not only 
is their argument transformed as $x \mapsto R^{-1}(\eta)x$ but they are also multiplied with an element of $\text{Spin}(2)$. In our representation, this is given by 
\begin{equation}
V(\eta) = e^{\frac{i\eta}{2} \sigma_z}
\end{equation}
 and can thus heuristically be understood as a generalized phase factor.
 Depending on whether they travelled above or below the cone, the respective localized parts of the wavefunction are multiplied with different generalized phase factors, according to $\eta = \alpha$ or $\eta = \beta$.

We note that with our coupled waveguide array tuned to host a conical geometry, we can realize such an interference experiment directly, since the intensity in the lattice can be used to infer the probability density, which transforms as a scalar under a general coordinate change \cite{Birell}. 
We now want to detail how to achieve this.

\subsection{Physical implementation in a waveguide array}
Our starting point is (\ref{DeString}), repeated here for convenience:
\begin{equation*}
i \partial_t \chi =i f\left( \sigma_y \partial_x + \sigma_x \partial_y\right) \chi+ m\sigma_z \chi
\end{equation*}
Here 
\begin{equation}
f = \frac1g = \left(\sqrt{x^2+y^2}\right)^{\Delta},
\end{equation}
and $\chi$ is a rescaled spinor, related to the true spinor as 
\begin{equation}\label{redef'}
\chi:= \sqrt{g} \psi = \frac1{\sqrt{f}} \psi.
\end{equation}
We discretize the space as
\begin{equation}\label{disc}
\begin{pmatrix}x\\y\end{pmatrix} \longmapsto \begin{pmatrix}\left(\frac12+n\right)\delta_x \\ \left( \frac12 +m\right)\delta_y \end{pmatrix} \ \ m,n \in \mathds{Z},
\end{equation}
which is useful because
\begin{equation}
\lim\limits_{x,y \rightarrow 0} g(x,y) \longrightarrow \infty,
\end{equation}
and we can cap this potentially singular behaviour in the field redefinition (\ref{redef'}) using (\ref{disc}) since no space point is directly placed in the center.
We define
\begin{equation}
\prescript{\ }{j}{\chi}_{n}^{m}:=\chi_j\left(\left( \frac12 + n\right) \delta_x ,\left( \frac12 + m\right)\delta_y\right),
\end{equation}
as well as
\begin{equation}
f(n,m):=   \left(\left( \left( \frac12 + n\right) \delta_x\right)^2+\left( \left( \frac12 + m\right) \delta_y\right)^2\right)^{ \frac{\Delta}2}, 
\end{equation}
\begin{equation}
\sqrt{g(n,m)} := \frac{1}{f(n,m)}
\end{equation}
and
\begin{equation}
h(a,b) := f(\lfloor a/2 \rfloor,b).
\end{equation}

We then take our waveguide lattice with three distinct modulations of the refractive index.
We assign rescaled spinor components $\prescript{\ }{j}{\chi}_{n}^{m}$ according to (\ref{prescription}), further rescaled by (\ref{Zfactor0}). 
To fix the values $\{Z_{mn}\}$, let us choose
\begin{equation}
Z_{00}=\frac{1}{\sqrt{f(0,0)}}.
\end{equation}
Then we have by (\ref{Zvalue})
\begin{equation}
Z_{mn}=\frac{1}{\sqrt{f(m,n)}} =\sqrt{g(n,m)}.
\end{equation}
Denoting
\begin{equation}
\left(g\prescript{\ }{j}{\psi}\right)_{n}^{m} := g(n,m) \cdot \prescript{\ }{j}{\psi}_{n}^{m},
\end{equation}
we are thus assigning rescaled spinor components according to:
\small
\begin{equation*}
\left[\begin{matrix}
-\left(g\prescript{\ }{1}{\psi}\right)_{n-1}^{m+2} & \left(g\prescript{\ }{2}{\psi}\right)_{n}^{m+2} & \left(g\prescript{\ }{1}{\psi}\right)_{n}^{m+2} &  -\left(g\prescript{\ }{2}{\psi}\right)_{n+1}^{m+2}\\

i\left(g\prescript{\ }{2}{\psi}\right)_{n-1}^{m+1} & -i\left(g\prescript{\ }{1}{\psi}\right)_{n}^{m+1} & -i\left(g\prescript{\ }{2}{\psi}\right)_{n}^{m+1}& i\left(g\prescript{\ }{1}{\psi}\right)_{n+1}^{m+1} \\

\left(g\prescript{\ }{1}{\psi}\right)_{n-1}^{m} & -\left(g\prescript{\ }{2}{\psi}\right)_{n}^{m} & -\left(g\prescript{\ }{1}{\psi}\right)_{n}^{m} & \left(g\prescript{\ }{2}{\psi}\right)_{n+1}^{m}  \\

-i\left(g\prescript{\ }{2}{\psi}\right)_{n-1}^{m-1} & i\left(g\prescript{\ }{1}{\psi}\right)_{n}^{m-1} & i\left(g\prescript{\ }{2}{\psi}\right)_{n}^{m-1} & -i\left(g\prescript{\ }{1}{\psi}\right)_{n+1}^{m-1}  \\
\end{matrix}\right]
\end{equation*}
\normalsize

To achieve the correct signs of the effective coupling constants we employ an amplitude prescription according to (\ref{signcoupling}) for the first modulation. 
The two remaining modulations are then used to achieve the necessary position dependent modulation of the absolute value of the coupling constants.
As detailed in (\ref{Aassignment}) and (\ref{Bassignment}), every site that has a non-zero amplitude for  one modulation has zero amplitude for the other and vice versa. Furthermore every non-zero amplitude of a given modulation is surrounded by sites that have zero amplitude for this modulation.
In the present case, the amplitudes on the lattice site $(ab)$ are then chosen such that 
\begin{equation}
J_0\left(\Xi_{ab}\right) = \sqrt{\frac{h(a,b)}{\max\limits_{n,m}f(n,m)}},
\end{equation}
where in the notation of section \ref{photlat}, $\Xi_{ab}=\frac{A_{ab}}{\alpha}$ or $\Xi_{ab}=\frac{B_{ab}}{\beta}$ depending on which amplitude is non-zero and we the maximum  is taken over all $\{(mn)\}$ considered in the actual setup.
This then yields the correct coupling constants for each wave guide $(a,b)$ to its neighbours to ensure the correct coupling of the spinor component $\prescript{\ }{j}{\Psi}_{ n}^{m}$ to its neighbours.

Next let us consider the input state preparation.
Given localized initial states $\Psi_0(y')$ and  $\Phi_0(y')$ in the coordinate system characterized by the angular variable $\theta$, we need to know how to express them in the global coordinate system characterized by $\tau$.
We know
\begin{equation}
(1-\Delta)\tau = \theta,
\end{equation}
and hence
\begin{equation}
[\tau - \theta]_{\text{mod}2\pi} = \Delta \tau.
\end{equation}
Since the vielbein is diagonal, the combined initial state $\Psi_\text{flat}(y') $ in the flat $\theta$-coordinate system then simply corresponds to the state
\begin{equation}\label{initialcont}
\Psi_{\rm initial}= V(\Delta\tau(x,y))\Psi_\text{flat}(y'(x,y)) .
\end{equation}
Here $\tau \equiv \tau(x,y)$ is given by the relation
\begin{equation}
(x,y)= \sqrt{x^2+y^2} \left(\cos(\tau(x,y)),\sin(\tau(x,y))\right),
\end{equation}
and (cf.~(\ref{COSMICSTRINGMETRIC}))
\begin{equation}
\begin{split}
&(y'^1,y'^2)(x,y)= \sqrt{(y'^1)^2 + (y'^2)^2} (\cos(\theta), \sin(\theta))\\
&=\frac{\sqrt{x^2+y^2}^{1-\Delta}}{(1-\Delta)} (\cos((1-\Delta)\tau), \sin((1-\Delta)\tau)).
\end{split}
\end{equation}

As input state in our lattice, we then prepare the rescaled discretized verion of of (\ref{initialcont}), given by
\begin{equation}
 \prescript{\ }{j}{\left(\Psi_\text{disc.}\right)}_{ n}^{m}:= g(n.m)   \prescript{\ }{j}{\left(\Psi_\text{initial}\right)}_{ n}^{m}.
\end{equation}
The evolution along the waveguides then mimics the time evolution of the approximated quantum system, i.e. that of a "particle" governed by the Dirac equation in conical space, in the global coordinate system. Hence if we want to know the probability density of the quantum  system at a given time, we can infer it from the intensity 
\begin{equation}
I_{ab}= |c_{ab}|^2
\end{equation}
 in our waveguide system at the corresponding transverse distance from the points where the input state was prepared.

To observe the "phase shift" of the gravitational Aharonov-Bohm effect, let us now prepare the input state
\begin{equation}
\Psi_\text{flat}(y') := \Psi_0(y')+ \Phi_0(y'),
\end{equation}
as discussed above in section \ref{extheo} (c.f.~also Figure \ref{initialstate}).
Note that the demands put on the  waveguide system that needs to be fabricated by the requirements on the position of the localised parts of the initial state can be mitigated by varying the discretization length or the physical (non-effective) coupling constants.  The requirement of the input state  being predominantly composed of modes corresponding to large enough momenta however, leads to a potentially complicated phase structure that might vary rapidly with the waveguide position for the input state. This might add additional challenges to the optical input state preparation.\\
After the input state is prepared and some "time" has passed, the probability density will then be given by (\ref{probdens}) with $z=z(x,y)$, since it simply transforms as a scalar under coordinate changes \cite{Birell}.
Note that 
\begin{equation}
\begin{split}
&(z^1,z^2)(x,y) = r(\cos(\phi),\sin(\phi))\\
&=\frac{\sqrt{x^2+y^2}^{1-\Delta}}{(1-\Delta)}(\cos(\phi(\tau(x,y))),\sin(\phi(\tau(x,y)))).
\end{split}
\end{equation}
The relation $\phi(\tau)$ between the angular coordinates is given by (\ref{rela}) and (\ref{relb}).
These equations imply 
\begin{equation}
\lim\limits_{\tau \downarrow 0}\phi(\tau) = (1-\Delta)\pi = \lim\limits_{\tau \uparrow 2\pi}\phi(\tau),
\end{equation}
and thus $\phi(\tau)$ is well defined.

In our waveguide array experiment, the amplitudes in the respective waveguides correspond to the spatially discretized version of the rescaled spinor components. Hence we can reconstruct the discretized probability density as 
\begin{equation}
\begin{split}
&\rho_{\Psi+\Phi}\left(t,z\left(\left[\frac12 + n \right]\delta_x,\left[\frac12 + m \right]\delta_y\right)\right)\\
&=\frac{1}{(g(n,m))^2} \left[I_{(2n)m} + I_{(2n+1)m} \right]\\
&= (f(n,m))^2 \left[I_{(2n)m} + I_{(2n+1)m}\right].
\end{split}
\end{equation}
Obtaining these experimentally and subsequently plotting them in the $z$ coordinate system (i.e.~as a function of $z$ and not as a function of $(x,y)$) using the relation $z(x,y)$, then precisely yields a discretized approximation of (\ref{probdens}).

Hence we can observe the deficit angle as the $\Delta$-dependent shift of this interference pattern. One might obtain the numerical value of $\Delta$ from the interference experiment described above by numerically calculating or obtaining from a separate waveguide experiment tuned to host flat space, the time evolution of the initial states $\Psi_0$, $\Phi_0$ in the flat coordinate system. Then the function
\begin{equation}
\begin{split}
	&f_\Delta(z):=  \bar \Psi\Psi(t,R_2(-(1-\Delta))z)\\
&	+\bar \Phi \Phi(t,R_2(-(1+\Delta)z)\\
	& + 2\Re\{\bar\Psi(t,R_2(-(1-\Delta))z) e^{i\sigma_z\Delta \pi}\Phi(t,R_2(-(1+\Delta))z)  \}.
\end{split}
\end{equation}
at fixed time $t$ can be fitted to the data obtained in the waveguide experiment detailed above. Here the fit proceeds with parameter $\Delta$. The resulting value of $\Delta$ yields the value of the "phase shift" $e^{i\Delta\pi}$.

\section{Conclusion}
We developed a scheme to implement the Dirac equation in $2+1$ dimensions in curved spacetime and classical electromagnetic background in a coupled waveguide array.
To do so, we had to introduce effective, site dependent coupling constants. This was achieved by periodically varying the refractive index of the respective waveguides. In the case of flat spacetime, one modulation sufficed, since only varying signs of the coupling constants needed to be achieved. For the general case, two more modulations and a spinor rescaling enabled the introduction of arbitrary site dependent couplings of spinor components. The electromagnetic potential was encoded into the refractive index too.
The implementation scheme in principle allows for implementations of many physically interesting settings. As an example, we devised a thought interference-experiment to observe a gravitational analogue of the Aharonov-Bohm effect in the spacetime of a cone. The global geometry of this locally flat setting introduced a generalized phase difference  between the time evolution of separately localized states.
Finally, we briefly explained how an analogue of this setting might in principle be implemented and how the associated generalized phase shift could be observed in a waveguide array.\\
\ \\
\textit{Acknowledgements:} 
C.K. acknowledges travel support provided by the Elite Master Program Theoretical and Mathematical Physics of the Free State of Bavaria, Germany. 
	D.A. was supported by the National Research Foundation, Prime
Minister’s Office, Singapore and the Ministry of Education, Singapore
under the Research Centres of Excellence programme. D.A. was also
partially funded by the Polisimulator project co-financed by Greece and
the EU Regional Development Fund, the European Research Council under
the European Union’s Seventh Framework Programme (FP7/2007–2013).

\appendix

\addcontentsline{toc}{section}{\bfseries Appendices}
\renewcommand{\thesubsection}{\arabic{subsection}}
\setcounter{subsection}{0}

\section*{Appendices}

\subsection{Introduction to Gravity in 2+1 dimensions}\label{gravity}
Since certain aspects are quite different from the more familiar 3+1 dimensional case, it is expedient to briefly discuss the most distinctive features of three dimensional gravity. We closely follow \cite{Vilenkin}.
\subsubsection{General remarks on 2+1 dimensional gravity}\label{generalgravity}
Our starting point is a three dimensional pseudo Riemannian manifold $(M,g)$. One can show that in 2+1 dimensions, the Ricci tensor completely determines the Riemann tensor \cite{Carlip}:
\begin{equation}
\begin{split}
R_{\mu\nu\rho\sigma} &= g_{\mu\rho} R_{\nu\sigma} + g_{\nu\sigma} R_{\mu\rho} - g_{\nu\rho} R_{\mu\sigma} - g_{\mu\sigma} R_{\nu\rho}\\
&-\frac12(g_{\mu\rho}g_{\nu\sigma}-g_{\mu\sigma}g_{\nu\rho})R.
\end{split}
\end{equation}
Gravity is defined via Einstein's field equations
\begin{equation}
G_{\mu\nu}=8\pi T_{\mu\nu},
\end{equation}
where we used the definition of the Einstein tensor,
\begin{equation}
G_{\mu\nu} := R_{\mu\nu} - \frac12 R g_{\mu\nu},
\end{equation}
and absorbed a possible cosmological constant into the Stress-energy tensor.
Note that throughout this paper we work in Planck units.
Since the vanishing of Einstein's Tensor implies a vanishing Ricci salar, it also implies that the Ricci tensor vanishes.
But in $2+1$ dimensions, this now immediately gives a vanishing Riemann tensor. 
There is thus no way that matter can communicate gravitationally in three dimensional spacetime and curvature at a specific point of the manifold is equivalent to matter being present at this point.
All effects that localized sources have are on the global geometry \cite{Deser}. 

Another interesting feature for this dimensionality is the fact that the Cotton tensor, given in coordinates by
\begin{equation}
C_{ijk}:=\nabla_k R_{ij}-\nabla{j}R_{ik}+ \frac14\left( \nabla_j Rg_{ik}   -\nabla_kRg_{ij} \right),
\end{equation}
plays the role of the Weyl tensor of higher dimensional pseudo Riemannian manifolds: In $2+1$ dimensions, the vanishing of the Cotton tensor is equivalent to conformal flatness, while in higher dimensions, conformal flatness is equivalent to the vanishing of the Weyl tensor.
It is not hard to see that a vanishing Ricci tensor implies a vanishing Cotton tensor \cite{Garcia}. Thus vacuum solutions to Einstein's field equations yield conformally flat metrics.
We now turn to a specific spacetime we want to study in greater detail.

\subsubsection{Example: The Cosmic string solution: Conical space}\label{topdef}
In $3+1$ dimensions, the simplest cosmic string solution is a solution that is invariant under translations in one spatial direction \cite{Vilenkin}. It is thus essentially a 2+1 dimensional system with a trivial dependence on a fourth spacetime coordinate, to be called z in what follows. For reasons of generality we first discuss it in the context of 3+1 dimensional gravity. Then we will discard the $z$-coordinate and consider the 2+1 dimensional system. 

During a symmetry breaking phase, for example in the very early universe, topological defects may arise naturally \cite{Sitenko2010,Isabelstring}.
Cosmic strings are such hypothetical, almost-one-dimensional topological defects in spacetime, which are analoguous to other linear defects that are familiar from condensed matter systems \cite{Isabelstring}.
The system of equations governing the dynamics of the gravitating string can be made tractable by assuming a vanishing string diameter, thus approximating it as a line of zero width through a distributional $\delta$-function energy-momentum tensor. Furthermore, one may assume a sufficiently weak gravitational field, so that the linearized Einstein equations provide a good approximation. For a detailed derivation, the reader is referred to \cite{Zwiebach} or \cite{Vilenkin} which we are following closely, below.

We consider a weak gravitational field in which the spacetime is almost Minkowskian,
\begin{equation}
g_{\mu \nu} = \eta_{\mu \nu} + h_{\mu \nu},
\end{equation}
with $ |{h_{\mu \nu}}| \ll 1 $. It is thus justified to linearize Einstein's equations in $h_{\mu \nu}$. Using the  harmonic gauge, specified by
\begin{equation}
\partial_\nu \left(h^{\nu}_{\mu} - \frac12 \delta^{\nu}_{\mu}  h^{\sigma}_{\sigma}\right) = 0,
\end{equation}
one may find the solution of Einstein's equations for a straight string along the z axis, for which
\begin{equation}
T^{\nu}_{\rho} = \mu \delta(x)\delta(y)\text{diag}(1,0,0,1) .
\end{equation}
The solution is found to be
\begin{align}
&h_{00}=h_{33}=0 \\
&h:= h_{11}=h_{22}=\mu \ln{\left(\frac{r}{r_0}\right)}
\end{align}
with $r:=\sqrt{x^2+y^2}$ and $r_0$ a constant of integration.

Consistency with the condition  $ |{h_{\mu \nu}}| \ll 1 $ then limits the domain of applicability of the weak field equations. However, the current coordinate system is not a suitable one for this purpose. Instead we go to the cylindrical coordinates, in which
\begin{equation}
ds^2=dt^2-dz^2-(1-h)(dr^2+r^2d\theta^2).
\end{equation}
Introducing 
\begin{equation}
(1-8\mu)\rho^2=(1-h)r^2
\end{equation}
one obtains to linear order in $\mu$,
\begin{equation}\label{almostconicalmetric}
ds^2 = dt^2-dz^2-d\rho^2 - (1-8\mu)\rho^2 d\theta^2.
\end{equation}
Introducing a new angular coordinate 
\begin{equation}
\tilde{\theta} = (1-4\mu)\theta
\end{equation}
the metric can be written in the Minkowskian form
\begin{equation}
ds^2 = dt^2-dz^2-d\rho^2 -\rho^2 d \tilde{\theta}^2
\end{equation}
with
\begin{equation}
0 < \tilde{\theta} < 2\pi (1-4\mu).
\end{equation}
again to linear order in $\mu$.

One therefore sees that to first order in $\mu$, the presence of the string introduces an azimuthal deficit angle $ 2 \pi \Delta$, where $\Delta \equiv 4  \mu$.
while the spacetime remains locally flat.
The surface of constant t and z thus has the geometry of a cone.
The weak field approximation (with $|h_{\mu\nu}| \ll 1$ ) is justified for $\mu \ll 1$. Observations constrain $G \mu \lesssim 10^{-5}$ for cosmological scenarios. 
The weak field metric   (\ref{almostconicalmetric}) coincides with
\begin{equation}\label{conicalmetricz}
ds^2 = dt^2-dz^2-d\rho^2 - (1-\Delta)^2\rho^2 d\theta^2
\end{equation}
to first order.
The validity of this metric can be extended beyond linear perturbation theory by taking the spacetime metric to be given by (\ref{conicalmetricz}) outside the domain of applicability of linear perturbation theory \cite{Vilenkin}. 

The metric (\ref{conicalmetricz}) is locally Minkowskian and is thus an exact solution of the vacuum Einstein equations.  However it induces  a non-trivial global geometry. 
We will now focus on the study of the 2+1 dimensional version of this metric, which is obtained by discarding the $dz^2$-term. 
At this point one might wonder which 2+1 dimensional object could play the role of a source for this metric.
It turns out that the metric for a spinning point source in 2+1 dimensional gravity and conformal coordinates, is given by \cite{Deser}
\begin{equation}\label{spinning}
ds^2= (dt + 4Jd\theta)^2 - R^{-8\mu}(dR^2+R^2 d\theta^2),
\end{equation}
with $\mu,J = const.$ and $\sqrt{g}T^{00}=\mu \delta(x_1)\delta(x_2)$, $T^{ij}=0$, $\sqrt{g}T^{0i}= \frac12J\epsilon^{ij}\partial_j \delta(x_1)\delta(x_2)$.
Here, $x_1,x_2$ are the cartesian coordinates corresponding to the cylindrical coordinate system the metric is expressed in.
Introducing a new radial coordinate 
\begin{equation}\label{COSMICSTRINGMETRIC}
r:= (1-4\mu)^{-1}R^{(1-4\mu)},
\end{equation}
this can be written as
\begin{equation}
ds^2= (dt + 4Jd\theta)^2 - dr^2 - (1-4\mu)^2r^2 d\theta^2.
\end{equation}
It is then easy to see that the metric is flat outside the source.  As we have discussed before, this has to be the case in 2+1 dimensions.  
The spacetime outside an arbitrary distribution of particles in the centre of mass frame confined to a region $r \textless r_0$ can be described by this metric for $r \textgreater r_0$  \cite{Vilenkin}. Within this region the curvature is not necessarily zero as there are sources present.

It is not hard to see that for  $J \textgreater 0$ the metric admits closed timelike curves (for example take $t:= t_0$, $r:= r_0 \textless 4GJ(1-4g\mu)$ and vary $\theta$ from $0$ to $2\pi$). To avoid this complication, we set $J:= 0$. Upon setting $\Delta:= 4\mu$ we are thus left with metric (\ref{conicalmetricz}) after discarding the z-direction:
\begin{equation}\label{conicalmetric}
ds^2 = dt^2-d\rho^2 - (1-\Delta)^2\rho^2 d\theta^2.
\end{equation}
This metric describes a conical space, which is a locally flat space with a wedge of angular size $2\pi\Delta$ removed and the two lines, along which the wedge was cut out, identified.
Note that local flatness means that each point has an open coordinate environment on which the metric in coordinates is the Minkowski metric. This is different from simply introducing normal coordinates, in which case the metric a priori only takes Minkowski form at a single point.
The metric in a global coordinate system can be inferred from (\ref{spinning}) with $J=0$
\begin{equation}\label{conspacemetric}
\begin{split}
ds^2& = dt^2 - R^{-2 \Delta}(dR^2+R^2 d\theta^2)\\
&=dt^2 - \left(\sqrt{x^2+y^2}\right)^{-2 \Delta}\left(dx^2+dy^2\right).
\end{split}
\end{equation} 
This is clearly of the form (\ref{prototype}) with 
\begin{equation}
e^\Psi = \left(\sqrt{x^2+y^2}\right)^{- \Delta}.
\end{equation}
Since the spacetime is locally flat, it is clear that a test particle initially at rest relative to the string will stay at rest and there is no Newtonian gravitational force.
The global geometrical difference with the Minkowski space however gives rise to interesting non-trivial effects, like for example double images \cite{Vilenkin}. 

The situation is reminiscent of the Aharonov-Bohm effect in electrodynamics. There, one considers a punctured plane and a vector potential $A$ that may be gauged away in any region with vanishing fundamental group, but not globally.
For a string, the curvature is confined to its core. We have chosen a vanishing diameter for simplicity, for the general case see \cite{Vilenkin}. The non-trivial global geometry however influences particles propagating in the flat region outside the core. A Minkowskian coordinate system can be chosen locally, but not globally.

\subsection{The Dirac equation in curved spacetime}\label{Diraccoordgeneral}
In local coordinates, the Dirac equation is given by \cite{Birell}
\begin{align}
\label{curveddiraceqn}
\left[ i\gamma^\mu\nabla_\mu -m\right]\psi(x) = 0.
\end{align}
Here we are using the vielbein $e\indices{^\mu_a}(x)$, defined by
\begin{equation}
e\indices{^\mu_a} e\indices{^\nu_b} \eta^{ab} = g^{\mu\nu},
\end{equation}
with the Minkowski metric $\eta$ to transform the local gamma matrices $\gamma^a$ according to
\begin{equation}
\gamma^\mu(x) = e\indices{^\mu_a}(x)\gamma^a.
\end{equation}
Note that $\{ \gamma^\mu(x),\gamma^\nu(x) \} = 2g^{\mu\nu}(x)$. 
The covariant derivative is given by
\begin{equation}
\nabla_\nu :=  \partial_\nu + \Omega_\nu + i A_{\nu},
\end{equation}
where the $A_{\nu}$ constitute the coordinate expression of the local $U(1)$ gauge connection. $\Omega_\nu$ is given by 
\begin{align}
\Omega_\nu(x) = -\frac{i}{4}\omega_{ab\nu}(x)\sigma^{ab}
\end{align}
where $\sigma^{bc} = i[\gamma^b,\gamma^c]/2$ and $\omega_{ab\nu}$ are the spin connection coefficients given by 
\begin{align}
\label{spinconnection}
\omega\indices{^a_b_\nu} = e\indices{^a_\mu} \partial_\nu (e\indices{^\mu_b}) + e\indices{^a_\mu}e\indices{^\sigma_b}\Gamma^\mu_{\sigma\nu}.
\end{align}
Note that $\omega_{ab\nu} = -\omega_{ba\nu}$.
For a pedagogical introduction the reader is referred to \cite{Koke}.

Specifically, for the metric given in (\ref{prototype}), the non-vanishing Christoffel symbols for the metric  are readily calculated to be
$\Gamma^0_{00}= \Phi_t$, $\Gamma^0_{10}= \Phi_x$ $\Gamma^0_{11}= \Psi_te^{2(\Psi-\Phi)}$, $\Gamma^0_{22}=\Psi_te^{2(\Psi-\Phi)}$, $\Gamma^1_{00}=\Phi_x e^{2(\Phi-\Psi)}$, $\Gamma^1_{10}=\Psi_x$, $\Gamma^1_{11}=\Psi_x$, $\Gamma^1_{12}=\Psi_y$, $\Gamma^1_{22}=- \Psi_x$, $\Gamma^2_{00}=- \Phi_y e^{2(\Phi - \Psi)}$, $\Gamma^2_{11}=-\Psi_y$, $\Gamma^2_{21}=\Psi_x$, $\Gamma^2_{22}=\Psi_y$. Here we denoted $f_\alpha :=\partial_\alpha f$ and omitted symbols that may be recovered from the given ones by index symmetry.
The vielbein $e\indices{^a_\mu}$ is readily calculated to be  $e\indices{^0_0} = e^\Phi$ $ e\indices{^1_1}= e\indices{^2_2}= e^\Psi $. 
Since the metric is diagonal, we have for the non-vanishing spin connection components that numerically
$\omega\indices{^a_b_c} =  \Gamma^a_{bc}$ if $a \neq b$.
From these, one obtains
\begin{align}\label{connectionmatrices}
\Omega_0 &= \frac14 \left(\omega_{010} [\gamma^0, \gamma^1] + \omega_{020} [\gamma^0, \gamma^2] \right) \nonumber \\
&=\frac14 \left(\Phi_xe^{\Phi-\Psi} [\gamma^0, \gamma^1] + \Phi_ye^{\Phi-\Psi} [\gamma^0, \gamma^2] \right), \nonumber \\
\Omega_1 &= \frac14 \left(\omega_{011} [\gamma^0, \gamma^1] + \omega_{211} [\gamma^2, \gamma^1] \right) \nonumber \\
&=\frac14 \left(\Psi_te^{\Psi-\Phi} [\gamma^0, \gamma^1] + \Psi_y [\gamma^2, \gamma^1] \right), \nonumber \\
\Omega_2 &= \frac14 \left(\omega_{022} [\gamma^0, \gamma^2] + \omega_{122} [\gamma^1, \gamma^2] \right)\nonumber \\
&=\frac14 \left(\Psi_te^{\Psi-\Phi} [\gamma^0, \gamma^1] + \Psi_x [\gamma^1, \gamma^2] \right).
\end{align}

Working with local gamma matrices $\{\gamma^a\}$, we may write down the Dirac equation as 
\begin{align}
&ie^{-\Phi}\partial_t\psi -\gamma^0 m\psi  + ie^{-\Psi}\gamma^0\gamma^1\partial_x \psi  + ie^{-\Psi}\gamma^0\gamma^2\partial_y \psi  \nonumber \\
&+i\left(e^{-\Psi}\gamma^0\gamma^1\frac{[\gamma^0,\gamma^1]}{4}+ e^{-\Psi}\gamma^0\gamma^2\frac{[\gamma^0,\gamma^2]}{4}\right)\Psi_te^{\Psi-\Phi}\psi   \nonumber \\
&+i\left( e^{-\Phi}\frac{[\gamma^0,\gamma^1]}{4} \Phi_x e^{\Phi-\Psi} + e^{-\Psi}\gamma^0 \gamma^2\frac{[\gamma^1,\gamma^2]}{4} \Psi_x  \right)\psi    \nonumber \\
&+i\left( e^{-\Phi}\frac{[\gamma^0,\gamma^2]}{4} \Phi_y e^{\Phi-\Psi} + e^{-\Psi}\gamma^0 \gamma^1\frac{[\gamma^2,\gamma^1]}{4} \Psi_y  \right)\psi  =0
\end{align}
Choosing $\gamma^0 = \sigma_z$, $\gamma^1=  - i\sigma_x$ and $\gamma^2=  i\sigma_y$, this equation becomes
\begin{equation}
\begin{split}
&i\left(\partial_t+\Psi_t\right)\psi= -i\sigma_ye^{\Phi-\Psi}\left(\partial_x + \frac{1}{2}\Psi_x +\frac{1}{2}\Phi_x \right)\psi\\
&-i\sigma_xe^{\Phi-\Psi}\left(\partial_y + \frac{1}{2}\Psi_y +\frac{1}{2}\Phi_y \right)\psi  +\sigma_ze^{\Phi} m \psi.
\end{split}
\end{equation}

Lastly, we remark that while it is well known that under a general diffeomorphism, a Dirac spinor can be taken to transform as a collection of  scalars \cite{Birell}, if we work in coordinates and apply  a change of coordinates from $\{x^\mu\}$ to $\{y^\mu\}$, then not only are we applying a diffeomorphism but we are also implicitly changing the frame in each tangent space with the help of which we express our Dirac spinor. This amounts to going from the basis $\{\partial/\partial x^\mu\}$ to the basis $\{\partial/\partial y^\mu\}$. In case the change can be represented by a proper Lorentz transformation in each tangent space, standard arguments yield that this can be lifted to the spinor representation as
\begin{equation}
\psi(x) \longmapsto \Lambda_\frac12(\omega_{ab}(\Lambda(x)))\Psi(\Lambda^{-1}(x)x),
\end{equation}
where $\Lambda$ is the corresponding matrix of the fundamental representation,
\begin{equation}\label{Localframechange}
\Lambda_\frac12(\omega_{ab}(\Lambda(x))) := \exp\left( \frac{i}{2} \omega_{ab}(x)\frac{i}{4}[\gamma^a,\gamma^b]\right).
\end{equation}
$\Lambda_\frac12(\cdot)$ denotes the spin $\frac12$ representation of the group element $\Lambda$ and the antisymmetric $\omega_{ab}$ characterizes our Lorentz transformation  \cite{Peskin}.

\subsection{Effective coupling constants: Justification for neglecting fast oscillating contributions}\label{besselapprox}
Our starting point is (\ref{effcoup}),
\begin{equation}
k_{[(a-1)\leftrightarrow a]b}^{\textit{eff}} :=\frac1T\int\limits_{z_0}^{z_0 +T} k_{[(a-1)\leftrightarrow a]b}(z)dz,
\end{equation}
where $T:= \frac{2\pi}{\omega}$, repeated here for convenience.
We assume that the scale of variation in the $z$-direction of the modes $c_{ab}(z)$ is much larger than $1/\alpha_j$ for all $j$. 
Furthermore we chose
\begin{equation}\label{natnum}
\alpha_j=q_j \omega,
\end{equation}
with $q_j$ positive natural numbers. 
We  expand the integrand using the generating function of the Bessel functions of the first kind
\begin{align}
\exp \left(i \sigma \sin(x) \right) = \sum\limits_{n\in \mathds{Z}} J_n(\sigma)e^{inx},
\end{align}
for each index $j$. 
To simplify notation, we write $A^j:=A^j_{(a-1)b}-A^j_{ab}$.
Interchanging summation and integration and using that 
\begin{align}
\frac1{2 \pi}\int_0^{2 \pi} \exp\left(i\sum\limits_jq_jn_j t \right)dt = \begin{cases}
1, & \text{if}\ \sum_jq_jn_j=0 \\
0, & \text{otherwise},
\end{cases}  
\end{align}
we see that
\begin{equation}
\begin{split}
k_{[(n-1)\leftrightarrow n]m}^{\textit{eff}} &= k_0  \prod_{j=1}^{\Lambda} J_0\left( \frac{ A^j}{2 \alpha_j}\right)\\
&+k_0\overbrace{\sum\limits_{\substack{\left(\sum\limits_{j=1}^\Lambda n_jq_j=0\right)\\ \textit{not all $n_j=0$  }}} \prod_{j=1}^{\Lambda} J_{n_j}\left( \frac{ A^j}{2 \alpha_j}\right)}^{=:C_\Lambda}.
\end{split}
\end{equation}
Since the convergence in (\ref{Bessel}) is absolute, we know that for fixed $x$
\begin{equation}
\sum\limits_{n \in \mathds{Z}} |J_n(x)| < \infty.
\end{equation}
Furthermore we note that $|J_{-n}(x)|=|J_{n}(x)|$. 
We now show by induction on $\Lambda$ that by appropriately choosing the modulation speeds $\alpha_j$, $C_\Lambda$ can be made arbitrarily small.
For this, we define
\begin{equation}
C_\Lambda^{|\cdot|}:= \sum\limits_{\substack{\left(\sum\limits_{j=1}^\Lambda n_jq_j=0\right)\\ \textit{not all $n_j=0$  }}} \bigg|\prod_{j=1}^{\Lambda} J_{n_j}\left( \frac{ A^j}{2 \alpha_j}\right)\bigg|.
\end{equation}
Clearly
$|C_\Lambda| \leq C_\Lambda^{|\cdot|}$
and it suffices to show that  $C_\Lambda^{|\cdot|}$ can be made arbitrarily small.
For this choose the natural numbers in (\ref{natnum}) as
\begin{equation}
q_j := 2^{j\Gamma},
\end{equation}
with a natural number $\Gamma$.
Since the case $\Lambda=1$ is trivial consider the case $\Lambda=2$ as the base case. There we have
\begin{equation}
C_2^{|\cdot|}=\sum\limits_{n_2\neq 0}\bigg|J_{-2^\Gamma n_2}\left(\frac{ A^1}{2 \alpha_1}\right) J_{ n_2}\left(\frac{ A^2}{2 \alpha_2}\right) \bigg|.
\end{equation}
Thus it follows that
\begin{align}
C_2^{|\cdot|}  & \leq   \left(\sum\limits_{n\neq 0} \bigg|  J_{ n}\left(\frac{ A^2}{2 \alpha_2}\right)  \bigg|   \right) \\
& \cdot \left(\sum\limits_{k\neq 0} \bigg|  J_{ (k 2^\Gamma )}\left(\frac{ A^1}{2 \alpha_1}\right)  \bigg|   \right) \stackrel{\Gamma \rightarrow \infty}{\longrightarrow} 0.
\end{align}
Hence we may make $C_2$ arbitrarily small by choosing $\Gamma$ large enough.

Now assume that $C_\Lambda^{|\cdot|}  \stackrel{\Gamma \rightarrow \infty}{\longrightarrow} 0$ for $\Lambda \leq p \in \mathds{N}$ and consider the case $\Lambda = p+1$.
To proceed, note that one may split the sum in $C_\Lambda^{|\cdot|}$ into a part where no $n_j =0$ and a part where always at least one $n_j=0$. The latter is then a sum of $\Lambda =p+1$ terms of the form  $|J_0(...)|  C_p^{|\cdot|}   $ with some of the summands in $C_p^{|\cdot|} $ set to zero. These terms have vanishing magnitudes by assumption and the remaining term is 
\begin{equation}
\sum\limits_{\substack{\left(\sum\limits_{j=1}^{p+1} n_j2^{j\Gamma}=0\right)\\ n_1,...,n_{p+1}\neq 0}} \bigg|\prod\limits_{j=1}^{p+1} J_{n_j}\left( \frac{ A^j}{2 \alpha_j}\right)\bigg| =: \tilde C^{| \cdot|}.
\end{equation}
We note that summands where we have  $|n_j| \leq 2^\Gamma - 1$ for all $1 \leq j \leq p+1$ do not occur, since in this case we have
\begin{equation}
\begin{split}
\bigg|\sum\limits_{j=1}^{p}n_j2^{j\Gamma}\bigg| &\leq (2^\Gamma-1)\sum\limits_{j=1}^{p}2^{j\Gamma}  \\
&=(2^\Gamma-1)\left(\frac{2^{\Gamma(p+1)}-1}{2^\Gamma - 1}\right)\\
&= 2^{\Gamma(p+1)}-1 < 2^{\Gamma(p+1)} \\
&< |n_{p+1}|2^{\Gamma(p+1)},
\end{split}
\end{equation}
violating the summation condition
\begin{equation}
\sum\limits_{j=1}^{p+1} n_j2^{j\Gamma}=0.
\end{equation}
Thus we in fact have
\begin{equation}
\tilde C^{| \cdot|} \equiv \sum\limits_{\substack{\left(\sum\limits_{j=1}^{p+1} n_j2^{j\Gamma}=0\right)\\\text{some} \ |n_j| > (2^\Gamma - 1)\\ n_1,...,n_{p+1}\neq 0}\\ }
\bigg|\prod\limits_{j=1}^{p+1} J_{n_j}\left( \frac{ A^j}{2 \alpha_j}\right)\bigg|.
\end{equation}
From this we infer
\begin{equation}
\tilde C^{| \cdot|} \leq \sum\limits_{k=1}^{p+1}  \sum\limits_{\substack{\left(\sum\limits_{j=1}^{p+1} n_j2^{j\Gamma}=0\right)\\ |n_k| > (2^\Gamma - 1)\\ n_1,...,n_{p+1}\neq 0}\\ }
\bigg|\prod\limits_{j=1}^{p+1} J_{n_j}\left( \frac{ A^j}{2 \alpha_j}\right)\bigg|.
\end{equation}
But then clearly (implicitly keeping the summation over all $n_j$  in the first row but not in the the second and third)
\begin{equation}
\begin{split}
\tilde C^{| \cdot|} \leq& \sum\limits_{k=1}^{p+1}  \sum\limits_{ |n_k| > (2^\Gamma - 1)}
\bigg|\prod\limits_{j=1}^{p+1} J_{n_j}\left( \frac{ A^j}{2 \alpha_j}\right)\bigg|\\
=&\sum\limits_{k=1}^{p+1} \left( \sum\limits_{ |n_k| > (2^\Gamma - 1)}
\bigg|J_{n_k}\left( \frac{ A^k}{2 \alpha_k}\right)\bigg|\right)\\
&\times \prod\limits_{j\neq k} \left( \sum\limits_{ n_j}
\bigg|J_{n_j}\left( \frac{ A^j}{2 \alpha_j}\right)\bigg|\right).\\
\end{split}
\end{equation}
This tends to zero as $\Gamma$ tends to infinity.

\subsection{Vanishing diagonal couplings}\label{diagonal}
In this appendix, we explain a possible method to realize vanishing diagonal couplings in a two dimensional square lattice. To avoid a cluttered notation, we will mainly resort to a pictorial representation. 
Our starting point are the approximate effective coupling constants (\ref{coupx}) and (\ref{coupy}) repeated here for convenience:
\begin{align}
k_{[(a-1)\leftrightarrow a]b} &\approx k_0  \prod_{j=1}^{\Lambda} J_0\left( \frac{ A^j_{(a-1)b}-A^j_{ab}}{2 \alpha^j}\right)\\
k_{a[(b-1)\leftrightarrow b]} &\approx k_0  \prod_{j=1}^{\Lambda} J_0\left( \frac{ A^j_{a(b-1)}-A^j_{ab}}{2 \alpha^j}\right)
\end{align}

We set $\Lambda =3$ and define $\chi:= \frac{A}{2 \alpha^1} = \frac{B}{2 \alpha^2} $ such that $J_0\left(\chi\right)=0$; E.g.~$\chi \approx 2.405$.
We then assign amplitudes according to
\begin{equation}
\left[\begin{matrix}
-A & \ & -\frac12A & \ &0 &\ -&\frac12A   \\
\ & \ & \   &\ & \  &\ & \      \\
\frac12 A & \ & A & \ &\frac12A &\ &0    \\

\end{matrix}\right],
\end{equation} 
for the first modulation and 
\begin{equation}
\left[\begin{matrix}
\frac12 B & \ & 0 & \ &\frac12B &\ &0    \\
\ & \ & \   &\ & \  &\ & \      \\
0 & \ & -\frac12B & \ &-B &\ &-\frac12B   \\
\end{matrix}\right],
\end{equation} 
for the second modulation, periodically extended through the lattice.
As is easily verified directly this implies that for the coupling between sites that are diagonally adjacent, there is always at least one Bessel function yielding the value zero.
The coupling in $y$-direction is so far then given by 
\begin{equation}
k_y = k_0 J_0\left(\frac32 \chi\right) J_0\left(\frac12 \chi\right).
\end{equation}
The coupling in $x$ direction is given by 
\begin{equation}
k_x=k_0 \left(J_0\left(\frac12\chi\right)\right)^2.
\end{equation}
Note that $k_y<0<k_x$ for $\xi \approx 2.405$.

To rectify this, we can use the third modulation.
We choose $\frac{C}{2\alpha_3} $ such that 
\begin{equation}
J_0\left(\frac{C}{2\alpha_3}\right) < 0,
\end{equation}
 and prescribe amplitudes according to
\begin{equation}
\left[\begin{matrix}
C & \ & C & \ &C &\ &C    \\
\ & \ & \   &\ & \  &\ & \      \\
0 & \ & 0 & \ &0 &\ &0  \\
\end{matrix}\right].
\end{equation} 

This then leaves $k_x$ unchanged and ensures $0<k_x,k_y$.
The remaining mismatch between the effective couplings can be absorbed into  different discretisation lengths in the two directions.

\end{document}